\documentstyle[aps]{revtex}
\def\gtsim{\vbox {\hbox{\lower 0.8\baselineskip \hbox{$>$}} \break
                 \hbox{\lower 0.2\baselineskip \hbox{$\sim$}} } }
\def\k{{\bf k}}
\def\p{{\bf p}}

\def\q{{\bf q}}

\def\R{{\bf R}}
\def\r{{\bf r}}

\def\vs{{\bf v}_s}
\def\v{{\bf v}}
\def\1v{{\bf v}}
\def\j{{\bf j}}
\def\A{{\bf A}}
\def\H{{\bf H}}
\def\K{{\cal K}}
\def\g{{\cal G}}
\def\F{{\cal F}}
\def\Z{{\cal Z}}
\def\qq{{\widehat Q}}

\begin{document}


\title{
Free Energy and Magnetic Penetration Depth 
of a $d$-Wave Superconductor\\ in the Meissner State }
\author{Mei-Rong Li,$^{a,b}$ P.J. Hirschfeld$^{c}$ and P. W\"olfle$^a$}

\address{
$^a$Institut f\"ur Theorie der Kondensierten Materie, Universit\"at
Karlsruhe, 76128 Karlsruhe, Germany\\
$^b$Department of Physics, Nanjing University, Nanjing 210093, P.R.China\\
$^c$Department of Physics, University of Florida, Gainesville, FL 32611,  USA \\ }

\maketitle

\begin{abstract}
We investigate the free energy and the penetration depth 
of a quasi-two-dimensional $d$-wave 
superconductor in the presence of a weak magnetic field by taking 
account of thermal, nonlocal, and nonlinear effects. In an approximation 
in which the superfluid velocity $v_s$ is assumed to be slowly varying, the free energy is
calculated and  compared with available results in several limiting cases. 
It is shown that either nonlocal or nonlinear effects
may cut off the linear-$T$ dependence of both the free energy and 
the penetration depth in all the experimental
geometries. At extremely low $T$,
the nonlocal effects will also generically modify the linear $H$ dependence
of the penetration depth (``nonlinear Meissner effect") in most experimental 
geometries, but for supercurrents oriented along the nodal directions, 
the effect may be recovered.   We compare our predictions with existing
experiments on the cuprate superconductors.

\end{abstract}


\section{INTRODUCTION}

Many unusual aspects of the unconventional
superconductivity in the cuprate and heavy fermion materials  are related to
the existence of nodes in the energy gap.  Although intensely discussed in
the 1980s and 1990s, most of these phenomena were already
known much earlier to students of the unconventional superfluid
$^3$He, which exhibits several phases, some with gap nodes and 
accompanying low-energy fermionic excitations.  For example,
power laws in specific heat, transport properties, and
NMR were predicted and measured in the anisotropic,
high-pressure "A" phase, along with 
singular responses to impurities, unusual vortex phases,
and even anomalous Josephson currents \cite{VollhardtWoelfle}. 
From this perspective, the one important difference between the
new superconductors and the $^3$He system is that they are
charged; unconventional  superconductivity can be uniquely
probed by studying the response of the system to an applied
electromagnetic field.  
\vskip .2cm
Even the simplest such phenomenon, the expulsion of a weak
applied magnetic field, or Meissner effect, is still the
subject of intense discussion today in the context of
unconventional superconductivity.  The new materials are
 strong type-II superconductors, with large values of
the London  penetration depth $\lambda_0$ and small coherence
lengths
$\xi_0$, such that the condition $\lambda_0\gg \xi_0$ is
expected to hold.  It was therefore anticipated in
the earliest theories of the Meissner effect in
unconventional superconductors\cite{Grossetal} that
the electrodynamics of the system could be treated as {\it
local}, i.e.,  as though the Cooper pairs were
 point objects,  neglecting the spatial variation of
the electromagnetic wave over the extent of the pair.
Under this rather plausible assumption, the simple 
anisotropic extension of the BCS theory is expected to hold,
\begin{equation}
{\lambda_0^2\over \lambda^2(T)}= 1-\left\langle{\hat k}_\parallel^2 \int d\xi_k
\left( -{\partial f
\over \partial E_k }\right)\right\rangle_{\rm FS},
\end{equation}
where the quasiparticle spectrum is
$E_k=\sqrt{\xi_k^2+\Delta_k^2}$,  $\xi_k$ is the normal-state electron band, 
$\Delta_k$ is the momentum-dependent order
parameter, $f$ is the Fermi function, ${\hat k}_\parallel$ is
the direction of the supercurrent, and $\langle\cdots \rangle_{\rm FS}$ 
represents a Fermi surface average.  The London penetration
depth at $T=0$ is given by
$\lambda_0\equiv\sqrt{mc^2/(4\pi n e^2)}$, and 
is typically one to several thousand Angstr\"om in the
new materials.
\vskip .2cm
At low temperatures, the local penetration depth calculated
from Eq. (1) may be shown\cite{Grossetal} to vary as a power law of
temperature,
$\lambda(T)-\lambda_0\sim (T/\Delta_0)^\alpha$, where
$\Delta_0$
is the gap maximum and the exponent $\alpha$ depends on
dimensionality, nodal topology, and the rate at which the gap
goes to zero near a nodal wave vector ${\bf k}_n$.  For
line nodes in a three-dimensional (3D) system (or point nodes in a 2D system)
like the $d_{x^2-y^2}$ state thought to characterize 
the cuprate superconductors, $\alpha$ is found to be 1.
Thus when Hardy {\it et al.} \cite{Hardyetal} measured a linear
temperature dependence in the YBa$_2$Cu$_3$O$_{6.95}$ (YBCO)
system down to a few degrees K, it was dramatic evidence in favor 
of line order parameter nodes and possible unconventional
pairing.
\vskip .2cm
Recently, Kosztin and Leggett\cite{KL} questioned
the theoretical basis for this identification, pointing out
that the approximation of  local electrodynamics used to derive
Eq. (1) is not valid for the quasiparticle states near the nodes
whose occupation determines the asymptotic low-temperature
behavior.  The coherence length $\xi_0\equiv v_F/{\pi\Delta_0}$
determines the scale of order-parameter variations, as in the
usual BCS theory, but an effective coherence length
$\xi_{0k}\equiv v_F/{\pi\Delta_k}$ appears in the
electromagnetic response, as noted in Ref. \cite{PHW}.
The divergence of $\xi_{0k}$ at the nodes means that quasiparticle 
states near the nodes must always be treated  nonlocally \cite{othernonloc}. 
A full nonlocal calculation gives a crossover from a $\delta\lambda\sim T$
regime for $E_{nonloc}\ll T\ll \Delta_0$ to a
$\delta\lambda\sim T^2$ for $T\ll E_{nonloc}$.  In the YBCO
system, this crossover scale $E_{nonloc}\simeq
\Delta_0\xi_0/\lambda_0$ is about 1 K, and thus the
identification of the $d$-wave state on the basis of the
local theory is not really in question; measurements on the
purest YBCO crystals down to 1 K or so have seen no deviation
from the linear behavior. \cite{newbonnhardypendepth}
Quadratic dependences $\delta\lambda\sim T^2$ measured in some samples 
are more likely due to impurity effects \cite{Grossetal,ProhammerCarbotte,felds}, which
also produce a $T\rightarrow T^2$ crossover.  
\vskip .2cm
An interesting perspective on the question of the linear $T$ dependence
of the penetration depth was provided recently by Schopohl and Dolgov \cite{SD},
who observed that {\it if} the
$d$-wave penetration depth temperature dependence were to remain linear down
to $T=0$, one would be unable to satisfy the third law of thermodynamics
(Nernst's theorem). In the framework of linear-response theory,
they expressed the change of the free energy in the presence of the magnetic
field (for fixed external current source) 
as a functional of the penetration depth, 
\begin{eqnarray}
\Delta{\cal F}^{(SD)}=-{1\over 8\pi} \int{d^3q\over (2\pi)^3}\left [
q^2+{1\over \lambda^2(\q,T)} \right ] |\A_\q(T)|^2,
\label{fesd}
 \end{eqnarray} 
where $\A_\q(T)$ is the Fourier component of the magnetic vector potential.
The resulting entropy is
\begin{eqnarray}
S^{(SD)}(T)=-{1\over 8 \pi} \int{d^3q\over (2\pi)^3}  {\partial
\over \partial T}\left[{1\over \lambda^2(\q,T)}  \right] |\A_\q(T)|^2. \label{entropysd}
\end{eqnarray}
The Nernst theorem requires $S$ to be 0 when $T\rightarrow 0$, inconsistent with
$\lambda(T)\sim T$.  
\vskip .2cm
Schopohl and Dolgov speculated that this result might imply the
instability of a pure $d$-wave state within BCS theory at sufficiently low temperatures.
It was pointed out in Ref. \cite{Volovik}, however, that the extra 
magnetic-field-induced quasiparticles that deplete the 
shielding current, leading to a magnetic-field-dependent penetration depth known 
as the nonlinear Meissner effect \cite{YipSauls,Xhuetal},
were neglected in this analysis.   Several authors have 
recently considered the effects of these nonlinear terms in the supercurrent on
transverse magnetization measurements\cite{ZuticValls} and on the structure of
the vortex lattice \cite{Aminetal,FGS}.
The basic idea is that, like impurity effects, 
the magnetic field itself may serve as a Cooper pair 
breaker that creates nodal quasiparticles, leading
to a temperature dependent penetration depth $\delta\lambda(T)\simeq T^2$ at temperatures 
below the  scale for nonlinear electrodynamics, $E_{nonlin}\simeq v_sk_F$ with 
$v_s$ a typical supercurrent velocity
and $k_F$ the Fermi wave vector.  The current authors\cite{LHWcomment} then provided a
synthesis of the various arguments given above, pointing out that
for any  experimental geometry with given disorder, external current distribution,
and Ginzburg-Landau parameter $\kappa$, a sample of
$d$-wave superconductor would inevitably 
avoid the violation of the Nernst theorem by creating a $T^2$
term in its penetration depth through a competition of nonlinear, nonlocal, and impurity
effects.
\vskip .2cm
Although this issue of principle has been resolved, an important prediction of the 
$d$-wave model for the electrodynamics of high-$T_c$ superconductors (HTS)
has not yet been
confirmed.  In the early discussion of the symmetry of the HTS order parameter,
Yip and Sauls proposed that the angular position of the gap nodes could be probed by a
measurement of the magnetic field dependence of the penetration depth. In the local limit and
for
$T\rightarrow 0$, they predicted that  the nonlinear effect induces a \textit{linear}-$H$ 
term in the
penetration depth, 
\begin{eqnarray}
{\delta \lambda (H)\over \lambda (0)}={\lambda(H)-\lambda(0)\over \lambda(0)}
\simeq 
\left \{
\begin{array}{ll}
H/H_0 & \;\;\;\;\;\;\;\; \v_s\parallel {\rm node} \\
{1\over\sqrt{2}} H/H_0 & \;\;\;\;\;\;\;\; \v_s\parallel {\rm antinode}
\end{array}
\right. \,, 
 \label{pdys}
\end{eqnarray}
where $H$ is the applied magnetic field at the surface and 
\begin{eqnarray}
H_0={3\Phi_0\over \pi^2 \xi_0\lambda_0} \label{h0}
\end{eqnarray}
is of order the thermodynamic critical 
field of the system with $\Phi_0$ the flux quantum.
 The prefactor of the linear $H$ term in Eq. (\ref{pdys}) 
depends generally on the angle $\theta$ that the supercurrent makes with
the crystalline $\hat{a}$ axis and is independent of $T$ (the $T$
dependence appears in the $H^2$ term as discussed recently
by Dahm and Scalapino \cite{Dahm}).
Several experimental groups tried to verify the prediction of
Yip and Sauls, but failed to identify a  linear $H$ term
which scaled  with the temperature according to the 
theory\cite{Maedaetal,CarringtonGiannetta,Bidinosti}.

There are three relevant energy scales in the low-energy 
sector in the Meissner state, $T$, $E_{nonlin}$, and $E_{nonloc}$. The 
first two are experimentally controlled parameters while the last
is an intrinsic one. Obviously, $\lambda(T)\sim T$ behavior should be 
expected at $T\gg E_{nonloc}, E_{nonlin}$. But in the opposite
limit, i.e., at extremely low $T$, either nonlinear or nonlocal effects 
may play a crucial role in modifying the linear $T$ behavior. Thus a full study
including both nonlinear and nonlocal effects 
in the Meissner state is necessary. A similar investigation 
has been performed by Amin, Affleck, and Franz \cite{Aminetal} in the mixed state 
where $H\geq H_{c1}$, the lower critical field. It was
found in this numerical study that the nonlinear and nonlocal corrections were of the
same order. This result is not surprising: in the mixed state the nonlinear energy
scale and the nonlocal one are not independent, since typical spatial variations take
place on a scale of the magnetic length or intervortex distance, fixed by the fluxoid
quantization. 

In this paper, we perform a full study of a quasi-two-dimensional (2D) $d$-wave
superconductor in the presence of a weak magnetic field. 
We start from a phenomenological BCS Hamiltonian 
with a quasiparticle energy spectrum
in the normal state, $\xi_\k=\epsilon_\k -\mu$ ($\mu$ is the chemical potential) and a
$d_{x^2-y^2}$-wave order parameter, $\Delta_\varphi=\Delta_0\cos 2\varphi$
($\varphi$ is the angle of $\k$ with the $\hat{a}$ axis).
Using the functional-integral approach, we calculate the free energy
and compare it with that obtained in other theories.
Our main approximation is that the supercurrent 
along the surface boundary is a slowly varying function of the 
distance to the surface. Thus
 we take $\v_s=(e/mc)\A_{\q=0}$ as the supercurrent 
 to which the quasiparticles couple, leading to the Doppler shift
 in the energy spectrum of particles and holes $\pm E_\k+\k\cdot \vs$,
 where $E_\k=[\xi^2_k+\Delta^2_\varphi]^{1/2}$. 
The response of the quasiparticles to the $\A_{\q\neq 0}$ modes 
is treated perturbatively, 
and the resulting response function involves both the nonlinear and nonlocal 
effects. From this response function we then calculate the 
penetration depth, and express it as a scaling function of the  two parameters
$E_{\rm nonloc}/T, \, E_{\rm nonlin}/T$. We show explicitly that, 
below a temperature $T^*$, 
the linear $T$ behavior of the penetration depth is always modified by
either nonlinear or nonlocal effects to a quadratic function of $T$, while 
the linear $H$ behavior in the penetration depth is also cut off 
by the nonlocal effects
when $H$ is below some characteristic critical field $H^\ast$, which is geometry
dependent. For most geometries
$H^\ast\simeq H_{c1}$, implying the unobservability of
the nonlinear Meissner effect, as discussed in our recent paper
on this subject \cite{LHWPRL}. But for some special 
geometries, in which the supercurrent is along nodal directions, 
the nonlinear Meissner effect may be recovered.
We also present a brief discussion on the geometry in 
which the magnetic field is along the conduction plane, which presents some special
features due to the highly layered nature of the cuprates. Finally, we investigate the
consequence of the spatially varying nature of the supercurrent and find that this
property does not influence the  qualitative behavior of the penetration depth.


The paper is organized as follows. In Sec. II, the model Hamiltonian and 
the partition function are obtained. In Sec. III, the expression of the 
free energy as well as its comparison with various theories in several
limiting cases is presented, and the supercurrent and the response function 
are discussed. Detailed study of the penetration depth in the presence
of a constant $v_s$ is given in Sec. IV.  Section V is devoted to 
concluding remarks and comparison with existing experiments. Finally, a discussion of the 
renormalization of the penetration depth due to  space-dependent $v_s$ 
is included in the Appendix.

\section{Model Hamiltonian and Partition Function}
We start from a phenomenological $d_{x^2-y^2}$-wave BCS mean-field Hamiltonian 
of a 2D system
 in the presence of a weak magnetic field \cite{note1}: 

\begin{eqnarray}
{\widehat H}_{\rm MF}&=&\sum_\sigma\int d^2\r c^\dagger_{\sigma}(\r)
\left \{ {1\over 2m} \left [-i{\vec \nabla} - {e\over c}\A(\r)\right ] ^2-\mu 
\right \} c_{\sigma}(\r)+
\int d^2\r d^2\r' \left [\Delta(\r,\r')c^\dagger_\uparrow(\r) c^\dagger_\downarrow
(\r')+ {\rm H.c.} \right ] \nonumber\\
&&+\int d^2\r d^2\r' V(\r-\r')|b(\r,\r')|^2,
\label{ham}
\end{eqnarray}
where 
$-V(\r-\r')<0$ is the effective attraction between electrons responsible for
the $d_{x^2-y^2}$ pairing, 
$\Delta (\r,\r')=-V(\r-\r')b(\r,\r')$ is the $d_{x^2-y^2}$-wave pairing order parameter
with $b(\r,\r')=\langle c_\downarrow (\r') c_\uparrow (\r)\rangle $, 
and $\A(\r)$ the magnetic vector potential.
It is well known that in the Meissner state the supercurrent formed by
the Cooper pairs effectively screens the magnetic field from penetrating into 
the bulk of the sample. Within the penetration depth away from the surface, 
both magnetic field and the supercurrent decay with the distance to the 
surface, and the pairing order parameter 
acquires, in principle, a spatially varying phase. 
As usual we write $\Delta(\r,\r')$ as
$\widetilde\Delta(\r-\r') e^{i\phi(\R)}$ with a real $\widetilde\Delta(\r-\r')$ and
$\R=(\r+\r')/2$. Note we neglect the supression of the magnitude $\widetilde\Delta $
at surface of the sample, since we expect these corrections to the penetration depth 
will be small in the limit $\lambda_0\gg \xi_0$ of interest.

The partition function of Hamiltonian (\ref{ham}) is 
\begin{eqnarray}
&&{\cal Z}= \Z_0 \int [Dd][Dd^*]\exp \left \{
-\int^{1/T}_0 d\tau \left [
\sum^2_{\alpha,\beta=1}\int d^2\r d^2\r' 
d^*_\alpha (\r) {\cal M}_{\alpha\beta} (\r,\r')d_\beta (\r')
 \right ] \right \} 
 \label{partitionold}
\end{eqnarray}
where
\begin{eqnarray}
\Z_0=\exp \left [- {1\over T} \left
(\frac{N_se^2}{2mc^2}\int d^2\r A^2(\r)
+{N_s\Delta^2_0\over V_0}\right ) \right] ,
\end{eqnarray}
with $N_s$ the total number of superconducting electrons,
$d_1(\r)=c_\uparrow (\r)$, $d_2(\r)=c^*_\downarrow (\r)$, and 
${\cal M}_{\alpha\beta} (\r,\r')$ is a matrix element of the $2\times 2$ matrix 
$\widehat{\cal M} $ in the Nambu representation,
\begin{eqnarray}
\widehat{\cal M}(\r,\r')=\left (
\begin{array}{cc}
\left \{ \partial_\tau+ {1\over 2m} \left [ i{\overrightarrow \nabla} 
+ {e\over c}\A(\r)\right ]^2
-\mu \right \}\delta(\r-\r')  &  \widetilde\Delta(\r-\r') e^{i\phi(\R)}   \\
\widetilde\Delta(\r-\r') e^{-i\phi(\R)}  &   
\left \{\partial_\tau- {1\over 2m}\left [ i\overleftarrow{\nabla} + {e\over c}\A(\r)
\right ]^2
+\mu \right \}\delta(\r'-\r) 
\end{array}
\right ) 
\end{eqnarray}

The use of a unitary transformation 
\begin{eqnarray}
{\widehat U}(\r,\r')=\left (
\begin{array}{cc}
e^{i\phi(\R)/2}  &  0\\ 
0 & e^{-i\phi(\R)/2}   
\end{array}
\right )
\end{eqnarray}
to $\widehat{\cal M}$ is made for eliminating the phase factor of $\Delta(\r,\r')$, 
leading to 
{\small
\begin{eqnarray}
&&\widehat{\widetilde{\cal M}}(\r,\r')=\widehat{U}^\dagger(\r,\r')
\widehat{\cal M}(\r,\r') \widehat{U}(\r,\r') \nonumber \\
&& = \;\;\; \left (
\begin{array}{cc}
\left \{ \partial_\tau+ \left [ {(i{\overrightarrow \nabla})^2\over 2m}-\mu \right ]
 + \v_s(\r) \cdot  i{\overrightarrow \nabla} + 
{1\over 2} m v^2_s(\r)  \right \}\delta(\r-\r')  &  \widetilde\Delta(\r-\r')  \\
\widetilde\Delta(\r-\r')  &   
\left \{ \partial_\tau - \left [ { ( i\overleftarrow{\nabla})^2\over 2m}-\mu 
\right ] -i\overleftarrow{\nabla}\cdot  \v_s(\r) - {1\over 2} m v^2_s(\r)  
 \right \}\delta(\r'-\r) 
\end{array}
\right ) ,
\label{Qtilde}
\end{eqnarray}
}
where 
\begin{eqnarray}
\v_s(\r)={e\A(\r)\over mc}-{\nabla\phi(\r)\over 2m} \label{vs1}
\end{eqnarray}
is the supercurrent velocity. In writing down Eq. (\ref{Qtilde}) we have used
the relation
${\overrightarrow \nabla} \cdot \v_s(\r)=0 $ which corresponds to the conservation of the supercurrent.  
$\v_s(\r)$ should be gauge invariant, which is guaranteed by the compensation
of the two terms in the right-hand side of Eq. (\ref{vs1}) under a gauge transformation. 
It is necessary to fix the gauge before proceeding. In the Meissner state the most
convenient gauge choice is the London gauge in which $\nabla\phi(\r)=0$, 
and hence
\begin{eqnarray}
\v_s(\r)={e\A(\r)\over mc}, \;\;\;\;\;\;\;  \nabla \cdot \A(\r)=0
\label{vsLondon}.
\end{eqnarray}

The partition function ${\cal Z}$ in Eq. (\ref{partitionold}) can be expressed 
in  momentum space as
\begin{eqnarray}
{\cal Z}={\cal Z}_0  
\exp \left [\sum_n {\rm tr} \; {\rm ln} \left ( \widehat{Q}^{(1)}(i\omega_n)
+\widehat{Q}^{(2)}
 \right ) \right ] ,
\label{pf3}
\end{eqnarray}
where $\omega_n $ is the fermion Matsubara frequency, 
\begin{eqnarray}
&& {\qq}^{(1)}_{\k,\k^\prime}(i\omega_n)=\left (\
\begin{array}{cc}
-i\omega_n+\xi_{\k}
+M_{\k,\k} &
\Delta_{\k}  \\
\Delta_{\k}  &   
-i\omega_n-\xi_{\k}
-M_{-\k,-\k}
\end{array}
\right )
\delta_{\k\k^\prime} =- \widehat{\cal G}^{-1}(i\omega_n,\k)\delta_{\k\k^\prime}
 , \label{matrix1} \\
&& \;\;\;\; {\qq}^{(2)}_{\k,\k^\prime}=\left (\
\begin{array}{cc}
M_{\k,\k^\prime}  & 0 \\
0 & -M_{-\k^\prime,-\k}
\end{array}
\right ). \label{matrix2}
\end{eqnarray}  
with $\Delta_\k=\int d^2\r \tilde{\Delta}(\r) \exp(i\k\cdot \r) 
\simeq \Delta_0\cos 2\varphi$  
on the Fermi surface, $M_{\k_1,\k_2}=-(e/mc)\k_2
\cdot \A_{\k_1-\k_2}+(e^2/ 2mc^2)\sum_\p\A_\p\cdot\A_{\k_1-\k_2-\p}$ 
with $\A_\k$ the Fourier component of $\delta\A(\r)=\A(\r)-\bar{\A}$ 
($\bar{\A}$ is the spatial average), $\xi_\k=\epsilon_\k-\mu$ 
is the energy spectrum of an electron in the normal state, and
$\widehat{\g}(i\omega_n,\k)$ is  Green's function 
matrix in Nambu's representation 
\begin{eqnarray}
\widehat{\g}(i\omega_n,\k)=-\left (\
\begin{array}{cc}
{i\omega_n+\xi_{\k}+M_{-\k,-\k}\over
W_\k}  & {\Delta_{\k}\over W_\k} \\
{\Delta_{\k}\over W_\k}  & {i\omega_n-\xi_{\k}
-M_{\k,\k}\over W_\k} 
\end{array}
\right ). 
\end{eqnarray} 
Here
\begin{eqnarray}
W_{\k}=-(i\omega_n+\k\cdot {\1v_s})^2+E^2_{\k} \,, \label{wk}
\end{eqnarray}  
where 
\begin{eqnarray}
\1v_s={e\bar{\A} \over mc}, \label {vs0} 
\end{eqnarray}
and $ E_\k=[(\xi_\k+(e^2/2mc^2)\int (d^2\p/(2\pi)^2)
A^2_\p)^2+\Delta^2_{\k}]^{1/2}$ .  
In Eq. (\ref{pf3}), we have explicitly separated the contributions to the free
energy from the homogeneous ($\bf q$=0) superflow and those corresponding
to ${\bf q}\ne 0$ components.

${\cal Z}$ in Eq. (\ref{pf3}) can be written as
\begin{eqnarray}
{\cal Z} &=& \Z_0 \exp \left \{ \sum_n {\rm tr} \; {\rm ln} \qq^{(1)}(i\omega_n)
\right \} \exp \left \{ \sum_n {\rm tr} \; {\rm ln} \left [
1-\widehat{\g}(i\omega_n)\qq^{(2)} \right ]  \right \} 
=\Z_0 \cdot \widetilde{\cal Z}_1\cdot\widetilde{\cal Z}_2 ,  \label{pf4} \\
\widetilde{\cal Z}_1&=&\exp\{\sum_n {\rm tr}\; {\rm ln} {\qq}^{(1)}(i\omega_n) \}
=\prod_\k\{1+e^{-(\k\cdot \v_s+E_k)/T} \}
\cdot\prod_\k\{1+e^{-(\k\cdot \v_s-E_k)/T} \}.    \label{pf5}
\end{eqnarray}
$\widetilde{\cal Z}_2$ may now be calculated by expanding in powers of  
$\A_{\q\neq {\bf 0}}$ . To second order, we have 
\begin{eqnarray}
{\rm ln}\widetilde{\cal Z}_2 &=& \sum_n{\rm tr} \; {\rm ln} 
(1-\widehat{\g}(i\omega_n)
{\qq}^{(2)}) \simeq \sum_n {\rm tr}\left [-\widehat{\g}(i\omega_n)
{\qq}^{(2)}-\frac{1}{2} \left (-\widehat{\g}(i\omega_n){\qq}^{(2)}\right )^2 \right ]
\nonumber\\
&=&-\frac{e^2}{2m^2c^2}\sum_{n} \int {d^2\k \over (2\pi)^2}
\int {d^2\q \over (2\pi)^2}
\left \{ \g_{11}(i\omega_n,\k_+) \g_{11}(i\omega_n,\k_-) 
(\k_-\cdot \A_\q) (\k_+\cdot \A_{-\q})  \right.\nonumber\\
&& +\g_{22}(i\omega_n,\k_+) \g_{22}(i\omega_n,\k_-) 
(\k_+\cdot \A_\q) (\k_-\cdot \A_{-\q}) 
+\g_{12}(i\omega_n,\k_+) \g_{21}(i\omega_n,\k_-)
(\k_+\cdot \A_\q) (\k_+\cdot \A_{-\q})
\nonumber\\
&&  \left.
+\g_{12}(i\omega_n,\k_+) \g_{21}(i\omega_n,\k_-)
(\k_-\cdot \A_\q) (\k_-\cdot \A_{-\q})    \right \},
\label{pf6}
\end{eqnarray}
where $\k_\pm =\k\pm \q/2$.
Note that only the paired electrons, i.e., the electrons near
the Fermi surface, can produce a diamagnetic effect. Thus we can replace $\k$
that couples to $\A_\q$ and $\v_s$ by $\k_F$. 
Adopting the London gauge $\q\cdot\A_\q=0$, 
we simplify ln$\widetilde\Z_2$ in Eq. (\ref{pf6}) as
\begin{eqnarray}
{\rm ln}\widetilde{\cal Z}_2\simeq 
-{e^2 \over m^2c^2}\sum_{n} \int {d^2\k \over (2\pi)^2} \int {d^2\q \over (2\pi)^2}
\frac{(i\omega_n+\k_F
\cdot\1v_s)^2+\xi_+\xi_-+\Delta_+\Delta_-}{W_+ W_-}
(\k_F\cdot \A_\q)(\k_F\cdot \A_{-\q}),
\label{pf7}
\end{eqnarray}
where $\xi_\pm=\xi_{\k_\pm}$, $\Delta_\pm=\Delta_{\k_\pm}$ and
$W_\pm=W_{\k_\pm}$.

\section{Free energy, supercurrent and response function}
The total free energy density of the superconductor in the Meissner state $\F$
is now given by the sum of contributions from both electrons and magnetic field, 
where the former 
can be calculated in terms of the partition function. We obtain
\begin{eqnarray}
\F=-T{\rm ln}\widehat{\Z}+{1\over 8\pi}\int d^2\r |\nabla \times\A(\r)|^2
=\F_1+\F_2 ,
\label{fe1}
\end{eqnarray}
where
\begin{eqnarray}
\F_1&=&\frac{1}{2}nmv^2_s-T\int {d^2\k\over (2\pi)^2} \left \{{\rm ln} 
\left [1+e^{-(\k_{\rm F}\cdot \1v_s+E_k)/T} \right ]
+{\rm ln} \left [1+e^{-(\k_{\rm F}\cdot \1v_s-E_k)/T} \right ] \right \}
+{n\Delta^2_0\over V_0},  \label{fe2}\\
\F_2&=&\frac{1}{2c}\int {d^2\q\over (2\pi)^2} {\cal K}(\q,\1v_s,T) A^2_\q
+\frac{1}{8\pi}\int {d^2\q\over (2\pi)^2} q^2 A^2_\q,  \label{fe3}
\end{eqnarray}
with 
\begin{eqnarray}
{\cal K}(\q,\1v_s,T)=\frac{ne^2}{mc} \left (
1+{2T\over nm}\sum_{n} \int{d^2\k \over (2\pi)^2} (\k_{\rm F})^2_\parallel
\frac{(i\omega_n+\k_F\cdot \1v_s)^2
+\xi_+\xi_-+\Delta_+\Delta_-}{W_+W_-}
\right )\, .   \label{rf1}
\end{eqnarray} 
$\F_1$ describes the kinetic energy of the quasiparticles 
with energy spectrum $\pm E_k$ shifted by $\k_F\cdot \1v_s$. 
Formally, it is nothing but the {\it Doppler shift} of the quasiparticle
with proper momentum $\k$ in the lab frame with local superfluid velocity
$\vs(\r)$. \cite{MakiTsuneto}  

It is interesting to look at $\F_2$ in Eq. (\ref{fe3}). In the linear limit
$\1v_s\rightarrow 0$, ${\cal K}(\q,\1v_s,T)$  in 
Eq. (\ref{rf1}) reduces to 
the usual linear response function of a superconductor in the Meissner 
state \cite{AGD}. The fact that ${\cal K}(\q,\1v_s,T)$
is indeed a response function can be seen from minimization of the free energy with 
respect to $\A_\q$, $\partial \F_2 / \partial \A_\q=0$, which gives the Fourier
component of the current inside the superconductor,   
\begin{eqnarray}
\j_{\rm SC} (\q)=-\K (\q,\1v_s,T) \A_\q, \;\;\;
{\rm for} \;\; \q\neq 0 . 
\label{current2}
\end{eqnarray} 
Therefore Eqs. (\ref{fe1})-(\ref{rf1})  describe
the Bogoliubov quasiparticles with the Doppler-shifted 
energy spectrum responding to a weakly spatially varying 
magnetic field. Note that $\K (\q,\1v_s,T)$ is obtained in an infinite sample. 
In the presence of a surface boundary, an external current $\j_{\rm ext} (\q) $ 
is introduced to coincide with the real boundary condition (see discussion in Sec. IV), 
leading to the total current as
\begin{eqnarray}
\j_{\rm tot} (\q)={c\over 4\pi} 
q^2 \A_\q=\j_{\rm ext}(\q)-\K (\q,\1v_s,T) \A_\q \,. 
\label{currenttot}
\end{eqnarray}


\begin{figure}[h]
\begin{picture}(180,180)
\leavevmode\centering\includegraphics{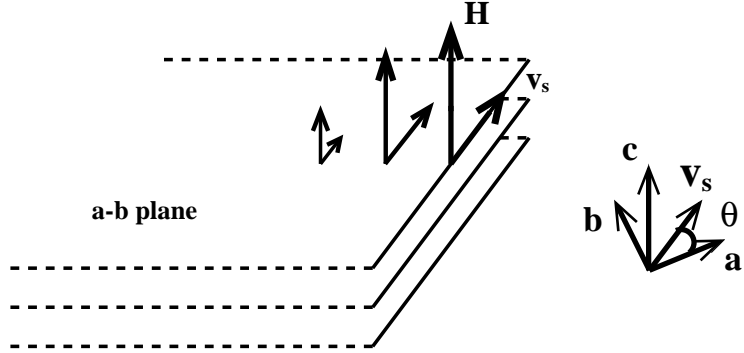}
\end{picture}
\caption{Geometry of (010) surface case.}
\end{figure}

By summing over the Matsubara frequencies, Eq. (\ref{rf1}) becomes
\begin{eqnarray}
&&\;\;\;\;\;\;\;\;\;\;\;\;\;\;\;\;
{\cal K}(\q,\1v_s,T)=\frac{c}{4\pi\lambda^2_0}
+\delta{\cal K}(\q,\1v_s,T),
\label{rf2} \\
&&\delta{\cal K}(\q,\1v_s,T)=-2{c\over 4\pi\lambda^2_0}
\left\langle ({\hat k}_F)^2_\parallel \int^\infty_0 d\omega {\rm Re}
\frac{[f(\omega-\1v_s\cdot\k_F)-f(-\omega-\1v_s\cdot\k_F)]\Delta^2_\k}
{\sqrt{\omega^2-\Delta_\k^2}
[\Delta_k^2+({\q\cdot\k_F/ (2m)})^2-\omega^2]}\right\rangle_{\rm FS} \\
&& \;\;\;\;\; =-\frac{c}{4\pi\lambda^2_0}\left \{
1-2 \left\langle ({\hat k}_F)^2_\parallel {\sinh^{-1}({\q\cdot \k_F\over 
2m\Delta_\k})
\over [{\q\cdot \k_F/ (2m\Delta_\k)}]\sqrt{1+[\q\cdot \k_F/ 
(2m\Delta_\k)]^2}}
\right\rangle_{\rm FS}  \right. \nonumber \\
&& \;\;\;\;\;\;\;\; \left. +2\left\langle ({\hat k}_F)^2_\parallel 
\int^\infty_0 d\omega {\rm Re}
\frac{[f(\omega-\k_F\cdot \1v_s)+f(\omega+\k_F\cdot \1v_s)]\Delta^2_\k}
{\sqrt{\omega^2-\Delta_\k^2}
\{\Delta_\k^2+[(\q\cdot \k_F)/(2m)]^2-\omega^2\} }\right\rangle_{\rm FS} 
\right \}
\label{rf3}
\end{eqnarray}  
where 
$f(x)$ is the Fermi function. 
$\delta{\cal K}(\q,\1v_s,T)$ is geometry dependent. The first two terms
in Eq. (\ref{rf3}) represent the nonlocal renormalization
of the $T=0$ response, while the third represents the combined nonlocal
and nonlinear corrections to the $T$ dependence. For a real 
quasi-2D system, like the high-$T_c$ compounds, with the $ab$ plane as the 
conduction plane, we first consider the geometry where the magnetic field 
is parallel to the $\hat{c}$ axis and thus $\1v_s$ and the penetration
direction $\q$ are in the $ab$ plane, and in general, $\1v_s$ makes an angle 
$\theta$ with the $\hat{a}$ axis. This geometry is shown in Fig. 1.  
Note that for $\v_s \not{ \parallel} $ antinode the possibility of the formation of an
Andreev bound state exists. Walter et al. \cite{walter} have argued that such 
states can have significant effects on the low-$T$ penetration depth for $\v_s \parallel $ node. 
The effects on the field dependence of the penetration depth 
are beyond the scope of the current formalism,
so results for these directions should be treated with caution. 
We will consider a different situation, where the magnetic field is in 
the $ab$ plane, at the end of Sec. IV. 

An examination of Eq. (\ref{rf3}) 
shows that the small regions around the four nodes give the main
contribution when $v_sk_F, T \ll \Delta_0$. This reflects the
low-energy quasiparticle excitations near nodes. It is straightforward to 
integrate out the angle of the Fermi wave vector, leading to
\begin{eqnarray}
\delta{\cal K}(\q,\v_s,T)=-\frac{c}{8\pi\lambda^2_0}\frac{T}{\Delta_0}
\sum_{l=\pm 1} u^2_{\theta l}
F_\lambda({\alpha u_{\theta, -l}\over T},{\varepsilon u_{\theta l}\over T}),
\label{rf4}
\end{eqnarray}  
where $\alpha=qk_F/(2\sqrt{2}m)$, $\varepsilon=v_sk_F/\sqrt{2}$,
$u_{\theta l}=|\cos\theta+l\sin\theta|,$ 
and $F_\lambda(z_1,z_2)$ is a two-parameter scaling function
\begin{eqnarray}
F_\lambda(z_1,z_2)=\frac{\pi}{4}z_1+\left [\ln (e^{z_2}+1)+\ln (e^{-z_2}+1) \right ]
-\int^{z_1}_{0}dx \left [{1\over e^{x-z_2}+1}+{1\over e^{x+z_2}+1} \right ]
\left [1-\left (\frac{x}{z_1}\right )^2 \right ]^{1/2}.
 \label{scalingf1}
\end{eqnarray}  
Now we may study two limiting cases.

\subsection{$v_sk_F \ll T$}

In this case we can expand $\F_1$ in Eq. (\ref{fe2}) up to the leading order of 
$v_sk_F/T$  
\begin{eqnarray}
\F_1\simeq \F^{(0)}_{\rm BCS}+
\left [
{1\over 2}nmv^2_s-{1 \over T}
\int {d^2\k\over (2\pi)^2} (\k_F\cdot\1v_s)^2
\frac{e^{E^{(0)}_\k/T}}{(1+e^{E^{(0)}_\k/T})^2} \right ],   \label{fe4}
\end{eqnarray}
where $E^{(0)}_k=(\xi^2_k+\Delta^2_k)^{1/2}$ and $\F^{(0)}_{\rm BCS}$ 
is the free energy density of the superconductor in the absence of a magnetic field,
\begin{eqnarray}
\F^{(0)}_{\rm BCS}=-T \int {d^2\k \over (2\pi)^2} \left [{\rm ln} (1+e^{-E^{(0)}_k/T})
+{\rm ln} (1+e^{E^{(0)}_k/T}) \right ]+{n\Delta^2_0\over V_0} \, . \label{fe5}  
\end{eqnarray}
Inserting Eq. (\ref{vs0}) into Eq. (\ref{fe4})  yields
\begin{eqnarray}
\F_1\simeq \F^{(0)}_{\rm BCS}+{1\over 2c}\K(0,0,T) A^2_0 \,, \label{fe6}
\end{eqnarray}
with $\K(0,0,T)={c\over 4\pi\lambda^2_0}+\delta\K(0,0,T)$ and
\begin{eqnarray}
\delta\K(0,0,T)=-{c \over 4\pi\lambda^2_0} (2\, {\rm ln} 2) \, {T\over \Delta_0} \, .
\label{rf5}
\end{eqnarray}
As for $\F_2$ in Eq. (\ref{fe3}), up to the leading order of $\A_\q$, the response function  
$\K(\q,\1v_s,T)$ is simply replaced by $\K(\q,0,T)$, the usual nonlocal response 
function \cite{AGD,KL}. It is straightforward to see from Eqs. (\ref{rf4})
and (\ref{scalingf1}) that
\begin{eqnarray}
\delta{\cal K}(\q,0,T)&=&\delta\K(0,0,T)
-\frac{c}{8\pi\lambda^2_0}\sum_{l=\pm 1}u^2_{\theta l}\left \{
\frac{\pi}{4}\frac{\alpha u_{\theta, -l}}{\Delta_0}
-\frac{1}{\Delta_0}\int^{\alpha u_{\theta, -l}}_0
dx f(x)\left [1-\frac{x^2}{\alpha^2 u_{\theta, -l}^2} \right ]^{1/2} 
\right \}    \\
&\simeq& 
\left \{
\begin{array}{ll}
-{c \over 4\pi\lambda^2_0} (2\, {\rm ln} 2) \, {T\over \Delta_0} , & \;\;\;\;\;\;\;
\alpha u_{\theta, -l} \ll T, \;\;\;\;\; l=\pm 1 \\
-{c \over 8\pi\lambda^2_0} \sum_{l=\pm 1}u^2_{\theta l} \left (
{\pi \over 4} {\alpha u_{\theta, -l} \over \Delta_0} + {3\over 2} \zeta(3) {T^3 \over \Delta_0 
\alpha^2 u_{\theta, -l}^2} \right ), & \;\;\;\;\;\;\;
\alpha u_{\theta, -l} \gg T, \;\;\;\;\; l=\pm 1   
\end{array} 
\right.  \, ,
\label{rf6}
\end{eqnarray}  
a special case of which, $\theta=0$, has been presented in Ref. \cite{KL}. 
In Eq. (\ref{rf6}) $\zeta(3)\simeq 1.20$ is Riemann's zeta function.

Therefore the change of the free energy due to the presence of the magnetic 
field is 
\begin{eqnarray}
\Delta \F=\F-\F^{(0)}_{\rm BCS}
=\frac{1}{8\pi}\int {d^2\q\over (2\pi)^2} [q^2+{4\pi\over c}{\cal K}^{(0)}(\q,0,T)] 
A^2_\q\; . \label{fe7}
\end{eqnarray}
Comparing Eq. (\ref{fe7}) with Eq. (\ref{fesd}), we
seem to find a sign discrepancy. However, notice that the free
energy in Eq. (\ref{fesd}) is calculated at {\it a fixed external current} and 
hence is the Gibbs free energy, whereas that in Eq. (\ref{fe7})
is the Helmholtz free energy, $\Delta\F^{({\rm Helm})}$,
since a {\it fixed} $\A_\q$ is assumed 
in the calculation. It can be checked that these two kinds of 
free energy are related to each other through the Legendre transformation
\begin{eqnarray}
\Delta\F^{({\rm SD})}=\Delta\F^{({\rm Helm})}-
{1\over c} \int {d^2\q\over (2\pi)^2} \j_{\rm ext}(\q)
\cdot \A_\q=-{1\over 8\pi}\int {d^2\q \over (2\pi)^2} 
[q^2+{4\pi\over c}{\cal K}^{(0)}(q,0,T)] A^2_\q\, . \label{fe8}
\end{eqnarray}
In the last equality, Eq. (\ref{currenttot}) was used.

The entropy can be calculated from the Helmholtz free energy in Eq. (\ref{fe7})
with $\A$ fixed, which is easily shown to be identical with that in Eq. 
(\ref{entropysd})
\begin{eqnarray}
\left. 
S(T)=-{\partial \Delta \F^{(Helm)} \over \partial T} \right |_{fixed \,\A}=S^{(SD)}(T) .
\label{entropy1}
\end{eqnarray}

Equations (\ref{fe7}) and (\ref{rf5}) imply that in the local limit
$\alpha\rightarrow 0$, $\Delta\F\propto T$.
If this linear $T$ behavior were to hold at $T\rightarrow 0$, the third 
law of thermodynamics, the Nernst's theorem, would be violated as pointed
out by Schopohl and Dolgov \cite{SD}. However, 
the intrinsic nonlocal effects may cut off  the linear $T$ 
term \cite{KL,LHWcomment}. Kosztin and Leggett showed, in the special
geometry where $\theta=0$, that the linear $T$ behavior of $\delta\K(\q,\v_s\rightarrow
0,T)$ will be changed to a higher power of $T$ when $T$ is smaller than $\alpha$.
In the case of more general $\theta$, the linear $T$ term is preserved until $T$ 
is smaller than both $\alpha u_{\theta l}, l=\pm 1$. For $\v_s$ along 
any node, $u_{\theta l}=0$ for $l=1$ or $-1$. The nonlocal effects thus
disappear along this nodal direction. In this case, the linear $T$ 
singularity may be cut off by another effect, the nonlinear effect,
which we now discuss.

\subsection{$v_sk_F \gg T$}

Notice that in Eq. (\ref{rf4}), $\varepsilon$ is always accompanied 
by $u_{\theta l}$, which, for $\theta$ near any node, may be very small. 
We consider first the situation where $\theta$ is not close to any node,
so that $\varepsilon u_{\theta l}\gg T$. 
The small perturbation parameter is now $T/(v_sk_Fu_{\theta l})$. 
The kinetic free energy $\F_1$ becomes
\begin{eqnarray}
\F_1={1\over 2}nmv^2_s-2T\int^\infty_{-\infty} dx N(x) {\rm ln} 
[1+e^{-x/T}]-\int {d^2\k \over (2\pi)^2} E_\k,  \label{fe9}
\end{eqnarray}
where
\begin{eqnarray}
N(x)=\int {d^2\k \over (2\pi)^2} \delta[x-(\1v_s\cdot\k_F+E_\k)]
={N_0\over 2\Delta_0} \sum_{l=\pm 1} (x+\varepsilon u_{\theta l}) \vartheta 
(x+\varepsilon u_{\theta l}), 
\label{dos}
\end{eqnarray}
with $\vartheta (y)$ the step function  
and $N_0$ the density of states on the Fermi surface in the normal state. 
Inserting Eq. (\ref{dos}) into Eq. (\ref{fe9}) we get
\begin{eqnarray}
\F_1=\F^{(0)}_{\rm BCS}+{1\over 2}nmv^2_s-{1\over 3}{\zeta_\theta\over \sqrt{2}} 
{nmk_F\over 2\Delta_0}v^3_s-{\pi^2\over 6} {nm\over \Delta_0 k_F} \sum_{l=\pm 1}
u_{\theta l} v_s T^2 ,
 \label{fe10}
\end{eqnarray}
where 
\begin{equation}
\zeta_\theta={1\over 2}\sum_{l=\pm 1}u^3_{\theta l}={1\over
2}\left(|\cos\theta+\sin\theta|^3+|\cos\theta-\sin\theta|^3\right) .
\end{equation}

The qualitative behavior of $\F_1$ as a function of both, $v_s$ and $T$, 
is consistent with that obtained by Volovik \cite{Volovik}.
Here we see that the prefactors of both the $v^3_s$ and $v_sT^2$ terms 
are $\theta$ dependent.

The scaling function $F_\lambda(z_1,z_2)$ in Eq. (\ref{scalingf1})
has the following asymptotic behavior at $z_2\gg 1 $: 
\begin{eqnarray}
F_\lambda(z_1,z_2)\simeq
\left \{
\begin{array}{ll} 
z_2\;\;\; &  z_1\ll z_2 \\
{\pi z_1\over 4}+{z_2(z^2_2+\pi^2)\over 6z^2_1} \;\;\; &  z_2\ll z_1\, .
\end{array}
\right. \, .
\label{scalingf2}
\end{eqnarray} 
This may be inserted into (\ref{rf4}) to find that 
$\delta\K(\q,\1v_s,T)=\sum_{l=\pm 1}\delta\K^{(l)}(\q,\1v_s,T)$, where
\begin{eqnarray}
\delta\K^{(l)}(\q,\v_s,T)=\left \{
\begin{array}{ll}
-{c\over 8\pi\lambda^2_0}u^3_{\theta l}{\varepsilon\over \Delta_0}
+{\cal O}(Te^{-\varepsilon/T}),  &
\alpha u_{\theta,-l}\ll \varepsilon u_{\theta l} \, , \\
-{c\over 8\pi\lambda^2_0}u^2_{\theta l}
[{u_{\theta,-l}\pi\over 4}{\alpha\over \Delta_0}
+{u^3_{\theta l}\over 6u^2_{\theta,-l}}{\varepsilon^3\over \alpha^2\Delta_0}
+{\pi^2u_{\theta l}\over 6u^2_{\theta,-l}}{\varepsilon T^2\over 
\alpha^2\Delta_0}] \, ,\;
& \alpha u_{\theta,-l}\gg \varepsilon u_{\theta l} \, .
\end{array}
\right.
\label{rf61}
\end{eqnarray}
It is clear from Eqs. (\ref{rf6}) and (\ref{rf61}), that the linear $T$ term in 
$\delta\K(\q,\1v_s,T)$, 
and hence in $\F_2$, is modified to a higher power of $T$ 
by either nonlinear or nonlocal effects. 
Combining this result with $\F_1$ in Eq. (\ref{fe9}), we see that a
thermodynamic instability, associated with the violation of 
the Nernst theorem, is avoided.

For $\theta$ near a nodal value, $u_{\theta l}\simeq 0$ for $l=1$ or $-1$. This 
means that 
those quasiparticles in nodal regions with wavevectors nearly perpendicular to 
$\v_s$ contribute negligibly to the nonlinear effect such that the linear 
$T$ singularity in the free energy may not be cut off.
These quasiparticles, however, acquire large nonlocal effects since $\q\perp\v_s$,
which may effectively modify the $T$ dependence of the free energy.
Similarly, it follows that the quasiparticles
which generate a small nonlocal effect produce much larger nonlinear effects.

With the expression for the free energy, we can easily obtain the supercurrent.
The part of the supercurrent arising from the kinetic free 
energy $\F_1$ in Eq. (\ref{fe10}) is given by
\begin{eqnarray}
\j^{(kin)}_{sc}=-{e\over m}{\partial \F_1 \over \partial \v_s}=-en\v_s +
en {v_s k_F\over 2\Delta_0} \left ( {\zeta_\theta \over \sqrt{2}} \v_s
+ {1\over 3\sqrt{2}} {\partial \zeta_\theta \over \partial \theta } 
\hat{z} \times \v_s \right ) ,
\end{eqnarray}
where $\hat{z}\parallel \hat{c}$.
It is clear that due to the nonlinear correction $\j^{(kin)}_{sc}$ is not  
parallel to $\v_s$ except for $\v_s \parallel$ node or antinode
in the case of which we recover the Yip-Sauls  
nonlinear supercurrent form \cite{YipSauls,Xhuetal},
\begin{eqnarray}
\j^{(kin)}_{sc}=\left \{
\begin{array}{ll}
-en\v_s(1-{v_sk_F\over 2\Delta_0}) , & \;\;\;\; \;\;\v_s \parallel 
{\rm node} \\
-en\v_s(1-{1\over\sqrt{2}}{v_sk_F\over 2\Delta_0}) , & \;\;\; \;\; \v_s \parallel 
{\rm antinode}
\end{array}
\right. \,. \label{current3}
\end{eqnarray}
The fact that  $\j^{(kin)}_{sc}$ is not  parallel to $\v_s$ for general $\theta$ 
leads to the interesting magnetic torque  phenomenon that has been discussed in 
Ref. \cite{Xhuetal}.

$\j(\q)$ can be obtained from Eqs. (\ref{current2}), (\ref{rf2}), and (\ref{rf61}).
It is nontrivial to examine the supercurrent in the local limit
\begin{eqnarray}
\j(\q\rightarrow 0)=
\left \{
\begin{array}{ll}
-{c\over 4\pi\lambda^2_0}(1-{v_sk_F\over \Delta_0})\A_{\q\rightarrow 0} , & 
\;\;\;\; \;\;\v_s \parallel {\rm node} \\
-{c\over 4\pi\lambda^2_0}(1-{1\over\sqrt{2}}{v_sk_F\over \Delta_0})\A_{\q\rightarrow 0} , & 
\;\;\; \;\; \v_s \parallel {\rm antinode} 
\end{array}
\right.  .
\label{current4}
\end{eqnarray} 
If we simply take $(e/mc)\A_{\q\rightarrow 0}$ 
as $\v_s$, we find that $\j(\q\rightarrow 0)$ is of 
the same structure as in $\j^{(kin)}_{sc}$ in Eq. (\ref{current3}), but with
an extra prefactor 2 in the nonlinear correction term. Note that 
$\j(\q=0)$ in the present theory is not equivalent to 
$\j^{(kin)}_{sc}$, but instead,
\begin{eqnarray}
\j(\q=0)=-{e\over m}{\partial \F \over \partial \v_s}
=\j^{(kin)}_{sc}-{e\over 2mc}\int {d^2\q \over (2\pi)^2} 
{\partial \K(\q,\1v_s,T) \over \partial \v_s} A^2_q. 
\label{current5}
\end{eqnarray} 
If $\K(\q,\v_s,T)$ can be replaced by its local-limit value, 
Eq. (\ref{current5}) becomes
\begin{eqnarray}
\j(\q=0)\simeq \left \{
\begin{array}{ll}
-{c\over 4\pi\lambda^2_0}(1-{v_sk_F\over \Delta_0}){mc\over e}\v_s, & 
\;\; \;\;\v_s \parallel {\rm node}, \\
-{c\over 4\pi\lambda^2_0}(1-{1\over\sqrt{2}}{v_sk_F\over \Delta_0})
{mc\over e}\v_s, & 
\;\; \;\; \v_s \parallel 
{\rm antinode}.
\end{array}
\right. 
\label{current6}
\end{eqnarray}

We conclude that the nonlinear correction we have obtained from the
$\q\rightarrow 0$ 
mode is formally two times that in Yip-Sauls theory. 
The reason for this discrepency is that we 
treat the response of the superconductor
to $\A_{\q\neq 0}$ modes perturbatively, while the nonlinear term 
($v^3_s$ term in the free energy) coming from the Doppler shift is not a 
perturbative result. To see this clearly, we make a naive perturbation
calculation of the response of the superconductor to a {\it constant} $v_s$.
The partition function is (Compare with  Eq. \ref{pf3})
\begin{eqnarray} 
\Z=\exp \left \{ \sum_n{\rm Tr} \,{\rm  ln} \left [\widehat{P}^{(1)}(i\omega_n)+\widehat{P}^{(2)}
\right ] \right \}
=\exp \left \{ \sum_n{\rm Tr} \,{\rm ln} \left [\widehat{P}^{(1)}(i\omega_n) \right ] \right \} 
\exp \left \{ \sum_n{\rm Tr} \,{\rm ln} \left [1+
\widehat{P}^{(1)-1}(i\omega_n) \widehat{P}^{(2)} \right ]  \right \}, \label{perturbpf}
\end{eqnarray} 
where
\begin{eqnarray}
\widehat{P}^{(1)}(i\omega_n) =
\left (
\begin{array}{cc}
-i\omega_n+\xi_\k  & \Delta_\k \\
\Delta_\k  & -i\omega_n-\xi_\k  
\end{array}
\right ) , \;\;\;
\widehat{P}^{(2)}=\k_F\cdot \1v_s
\left (
\begin{array}{cc}
1  & 0 \\
0  & 1  
\end{array}
\right ) ,  \label{perturbmatrix}
\end{eqnarray} 
with $\widehat{P}^{(2)}$ assumed small.  The free energy is
\begin{eqnarray}
\F=-T{\rm ln} \Z= -T\sum_n {\rm Tr} \,{\rm ln} [\widehat{P}^{(1)}(i\omega_n) ] 
-T \sum_n{\rm Tr} \,{\rm ln} [1+
\widehat{P}^{(1)-1}(i\omega_n)\widehat{P}^{(2)}].  \label{perturbfe}
\end{eqnarray} 
The $v^3_s$ term  of present interest is expected to arise from 
\begin{eqnarray}
-T\sum_n {\rm Tr} \left \{ {1\over 3} \left [\widehat{P}^{(1)-1}(i\omega_n)
\widehat{P}^{(2)}   \right ]^3 \right \}
=-{1\over 3} T \sum_n {\rm Tr}(\1v_s\cdot \k_F)^3 
\left (
\begin{array}{cc}
{-i\omega_n-\xi_\k \over (i\omega_n)^2-\xi^2_\k-\Delta^2_\k}  & 
{-\Delta_\k \over (i\omega_n)^2-\xi^2_\k-\Delta^2_\k} \\
{-\Delta_\k \over (i\omega_n)^2-\xi^2_\k-\Delta^2_\k} &
{-i\omega_n+\xi_\k \over (i\omega_n)^2-\xi^2_\k-\Delta^2_\k}
\end{array}
\right ) ^3 =0.
\end{eqnarray} 
A nonlinear term $\propto v^3_s$ cannot be obtained from the perturbation theory,
since the perturbation calculation implies $\1v_s\cdot \k_F\ll 
\Delta_\k$, while the nonlinear $v^3_s$ term comes from those $\k$ points near 
nodes on the Fermi surface satisfying  $\Delta_\k<\v_s\cdot \k_F$. Therefore
the crucial step of taking $(e/mc)A_{\q\rightarrow 0}$ as $\1v_s$ in 
obtaining Eq. (\ref{current6}) is an approximation since the 
couplings of the quasiparticles to $\A_{\q\neq 0}$ are treated perturbatively.
 Nevertheless we expect that the qualitative
scaling behavior will be correctly reproduced with the
present formalism. The renormalizations of the naive results to be expected from 
treating the magnetic field in a fully self-consistent manner are discussed in
the Appendix.

\section{penetration depth}
The penetration depth for a half infinite system may be defined as
\begin{eqnarray}
\lambda={1\over H}\int^\infty_0 H(y) dy, \label{pd1}
\end{eqnarray} 
where $H$ is the magnetic field at the surface. A {\it specular} scattering 
surface boundary condition on the 
quasiparticle wave functions is assumed, which replaces
the surface with a current sheet of the form
\begin{equation}
\j_{ext}(y)=-{c\over 2\pi}H\delta(y)\hat{x}. \label{currentext}
\end{equation}
in an infinite system \cite{Tinkham1}.

\subsection{Penetration depth at a constant $v_s$}
As a first step, in this subsection, we study the penetration depth 
in the system characterized by an electromagnetic response function calculated in
the presence of  a
constant
$v_s$.   Since the true, self-consistently determined superfluid velocity 
decays with distance from the surface, this procedure introduces
errors in the nonlinear terms, which we discuss below and estimate in some detail in
the Appendix.  For the moment, we consider Maxwell's Eq. 
\begin{equation}
\nabla^2\A=-{4\pi\over c} (\j_{ext}+\j_{sc}) \label{ampere}
\end{equation}
and the London Eq. (\ref{current2}), obtaining
\begin{eqnarray}
\lambda_{\rm spec}=\frac{2}{\pi}\int^\infty_0 \frac{dq}{4\pi{\cal K}
(\q,\v_s,T)/c+q^2}\simeq\lambda_0-{8\over c}\int^\infty_0 dq\frac{\delta{\cal K}
(\q,\v_s,T)}
{(\lambda^{-2}_0+q^2)^2}.
\label{pd2}
\end{eqnarray} 
Here $\v_s$ is now to be understood as its value at the surface 
$v_s=v_s(y=0)=e\lambda_0H/mc$
as given by the solution to the linear, local electrodynamics problem. 
Obviously, $\lambda_{\rm spec}$ includes 
both the nonlocal and nonlinear effects. It is expected to reduce exactly 
to the nonlocal expression of Kosztin and Leggett\cite{KL} if 
the $\v_s$ dependence 
is neglected, and (qualitatively) to the nonlinear expression 
of Yip and Sauls
\cite{YipSauls} if the $q$ dependence is neglected. We first go over 
these two limiting cases.

For the linear limit $v_s\rightarrow 0$, the qualitative behavior of the penetration 
depth depends on two effective nonlocal energy scales, 
$E^{(+)}_{\rm nonloc}=v_Fu_{\theta l_1}/\lambda_0$ 
and $E^{(-)}_{\rm nonloc}=v_Fu_{\theta l_2}/\lambda_0$ 
for $l_1,l_2=\pm 1$ and $u_{\theta l_1}\geq u_{\theta l_2}$.  
It is shown that 
\begin{eqnarray}
{\delta\lambda ^{\rm (lin)}_{\rm spec}\over \lambda_0}\simeq
\left \{
\begin{array}{ll}
{1\over 2} ({\rm ln}2)\sum_{l=\pm 1}u^2_{\theta l} {T\over \Delta_0}, 
\;\;\; & {\rm for} \;\; E^{(+)}_{\rm nonloc},
E^{(-)}_{\rm nonloc}\ll T\\
{1\over 2} ({\rm ln}2) u^2_{\theta l_1} {T\over \Delta_0}+
{\pi\over 16\sqrt{2}} \kappa^{-1} u_{\theta l_1}u^2_{\theta l_2}
, \;\;\; & {\rm for} \;\; 
E^{(-)}_{\rm nonloc}
\ll T\ll E^{(+)}_{\rm nonloc}\\
{\pi\over 16\sqrt{2}} \kappa^{-1}\sum_{l=\pm 1} u_{\theta,-l}u^2_{\theta l}
+0.80 \, \kappa \,({T\over \Delta_0})^2 \sum_{l=\pm 1} {u^2_{\theta l}\over u_{\theta,-l}}
, \;\;\; & {\rm for} \;\; 
T\ll E^{(+)}_{\rm nonloc}, E^{(-)}_{\rm nonloc} 
\end{array}
\right. . \label{pd3}
\end{eqnarray} 
For $\theta \sim 0$, 
$E^{(-)}_{\rm nonloc}\simeq E^{(+)}_{\rm nonloc}$ and there is no intermediate
parameter region $E^{(-)}_{\rm nonloc}\ll T\ll E^{(+)}_{\rm nonloc}$. In 
this case, we recover Kosztin and Leggett's result \cite{KL}. However,
for $\theta$ near any node, $E^{(-)}_{\rm nonloc}$ disappears, and 
the linear $T$ behavior of the penetration depth is 
preserved even at $T\ll v_F/\lambda_0$, but with a reduced prefactor. 
We conclude that in the linear limit the nonlocal effects fail to cut off 
the linear $T$ dependence of the penetration depth when $v_s$ is along 
nodal directions. On the other hand, it is precisely in this limit that the 
effects of Andreev bound states on the penetration depth are expected to be largest. 
Although the influence of these states on the field dependence has not
yet been calculated, it appears unlikely that the naive Yip-Sauls result can 
apply.

For the local limit $\q\rightarrow 0$, Eq. ({\ref{pd2}) becomes
\begin{eqnarray}
\lambda^{({\rm loc})}_{\rm spec}=\left (
\frac{c}{4\pi {\cal K}(\q\rightarrow 0,\v_s,T)} \right ) ^{1/2}.
\label{pd4}
\end{eqnarray}
Again, for a general $\theta$, there are two effective nonlinear
energy scales, $E^{(+)}_{\rm nonlin}=v_sk_Fu_{\theta l_1}$
and $E^{(-)}_{\rm nonlin}=v_sk_Fu_{\theta l_2}$.
One gets
\begin{eqnarray}
\frac{\delta\lambda^{({\rm loc})}_{\rm spec}}{\lambda_0}
\simeq
\left \{
\begin{array}{ll}
{1\over 2} ({\rm ln}2)\sum_{l=\pm 1}u^2_{\theta l} {T\over \Delta_0},
& \;\;\; \;\; E^{(+)}_{\rm nonlin},E^{(-)}_{\rm nonlin}\ll T \\
{1\over 2}({\rm ln}2)u^2_{\theta l_2} {T\over \Delta_0}+
{u^3_{\theta l_1}\over 2\sqrt{2}}\frac{v_sk_F}{2\Delta_0} , 
& \;\;\;  \;\; E^{(-)}_{\rm nonlin}\ll T\ll E^{(+)}_{\rm nonlin} \\
{\zeta_\theta \over \sqrt{2}}\frac{v_sk_F}{2\Delta_0}
+{\cal O}(Te^{-v_sk_F/\sqrt{2}T}),
& \;\;\;  \;\; T\ll E^{(+)}_{\rm nonlin},E^{(-)}_{\rm nonlin}
\end{array}
\right. \,.
\label{pd5}
\end{eqnarray}  

Now we are in the position to study the penetration depth with 
both nonlocal and nonlinear effects. We are basically interested in the case of 
$T\ll v_sk_F$. Inserting Eqs. (\ref{rf4}) and (\ref{scalingf2}) into 
Eq. (\ref{pd2}) leads to 
\begin{eqnarray}
{\delta\lambda_{\rm spec}\over \lambda_0}=\sum_{l=\pm 1}
{\delta\lambda^{(l)}_{\rm spec}\over \lambda_0},
\label{pd6}
\end{eqnarray} 
where
\begin{eqnarray}
\frac{\delta\lambda^{(l)}_{\rm spec}}{\lambda_0}
\simeq\frac{k_F u^3_{\theta,-l}}{8\sqrt{2}\pi m^3\lambda^3_0\Delta_0v^2_s}\eta_1(h_{\theta l})
+\frac{T^2 u^3_{\theta,-l}}{4\sqrt{2}\pi m^3\lambda^3_0\Delta_0v^4_sk_F u^2_{\theta l}}
\eta_2(h_{\theta l}), 
\label{pd7}
\end{eqnarray} 
with
\begin{eqnarray}
\eta_1(x)&=&4x^3 \left [\frac{2x}{1+4x^2}+\arctan (2x) \right ]+\frac{x^2 (3\pi+16 x^2+96x^4
-48x^3(1+4 x^2)\arctan (1/2x)}{6(1+4x^2)}, \nonumber\\
\eta_2(x)&=&\frac{4\pi^2 x^4[2+12 x^2-6x(1+4 x^2)\arctan(1/2x)]}{3(1+4x^2)},
\label{eta12}
\end{eqnarray} 
and 
\begin{eqnarray}
h_{\theta l}={u_{\theta l}\over u_{\theta,-l}}h \,, \;\;\;\;\;\;\;\;\;\;\;\;
h=m\lambda_0v_s={3\over \pi}\kappa {H\over H_0} \,, \label{hh}
\end{eqnarray} 
defining the competition between the nonlinear and nonlocal effects, 
with $H_0$ as defined in Eq. (\ref{h0}). 
It is straightforward to get the asymptotic behavior of the penetration depth:
\begin{eqnarray}
\frac{\delta\lambda^{(l)}_{\rm spec}}{\lambda_0}\simeq
{\pi\over 4\sqrt{2}}u^3_{\theta l}\kappa^{-1} h+c_{\theta l1}
\kappa^{-1}\frac{1}{h^2}
+c_{\theta l2}\kappa{1\over h^4}{T^2\over \Delta_0^2} , \;\;\;\;
\;\;\; h_{\theta l} \gg 1, 
\label{pd8} 
\end{eqnarray}
where the nonlinear effect dominates, with $c_{\theta l1}=0.0082 u^3_{\theta,-l}$ 
and $c_{\theta l2}=0.0059 u^3_{\theta,-l}/u^2_{\theta l}$, and 
\begin{eqnarray}
\frac{\delta\lambda^{(l)}_{\rm spec}}{\lambda_0}\simeq
{\pi\over 8\sqrt{2}}d_{\theta l1}\kappa^{-1}
+d_{\theta l2}\kappa^{-1} h^2 + 
d_{\theta l3} \kappa {T^2\over\Delta_0^2} , \;\;\;\;
\;\;\; h_{\theta l} \ll 1,
\label{pd9}
\end{eqnarray} 
where the nonlocal effects dominate, 
with $d_{\theta l1}=0.50 u^2_{\theta l}u_{\theta,-l}$, $d_{\theta l2}=1.09 
u^4_{\theta l}/u_{\theta,-l}$, and $d_{\theta l3}=0.47 u^2_{\theta l}/u_{\theta,-l}$. 
  In the local, $H\rightarrow 0$ limit, the result reduces to
\begin{equation}
{\delta\lambda_{spec}\over \lambda_0} \simeq {3 \over 2\sqrt{2}} 
\zeta_\theta {H\over H_0}, \label{pdlocal}
\end{equation}
which is apparently a factor of 3/2 larger than the result (\ref{pdys}).
In fact, the comparison is a bit more subtle, as we have taken a definition
of the penetration depth which differs slightly from that of Yip and Sauls. 
\cite{YipSauls} This comparison is discussed further in the Appendix.

 From Eqs. (\ref{pd8}) and (\ref{pd9}) one now sees clearly that the 
 linear $H$ dependence of $\delta\lambda^{(l)}_{\rm spec}$ 
at $h_{\theta l}\gg 1$
is modified to an $H^2$ dependence when $h_{\theta l}\ll 1$, implying that 
the nonlinear effect may be cutoff by the nonlocal effects. 
The crossover field for this to happen is defined by $h_{\theta l}\sim 1$.
For most $\theta$, $u_{\theta l}$ and $u_{\theta,-l}$ are order of unity 
and the crossover field is simply defined by $h\sim 1$, i.e., 
$H^\ast\simeq \pi\kappa^{-1}H_0/3
\simeq H_{c1}$, the lower critical field. Since at $H\geq H_{c1}$ the Meissner 
state is unstable to the Abrikosov vortex state, the nonlinear Meissner
effect is effectively unobservable due to the nonlocal effect 
in this geometry \cite{LHWPRL}. In Fig. 2, we display the magnetic field 
dependence of the penetration depth for $\theta=0$. 
It is clear that in this case the predicted linear behavior in $H$ is only 
recovered well above $H_{c1}$. On the other hand, for $\theta$ 
very close to a node, min($h_{\theta l})\sim 0$, meaning that the crossover 
field can be so small that the 
nonlinear effect cannot be cut off by the nonlocal one \cite{LHWPRL}. The physical
interpretation of this is that only those 
quasiparticle excitations near node regions can reduce the shielding current. When 
$\v_s$ is along a nodal direction, those quasiparticles near this node acquire a large
nonlinear energy shift, but on the contrary a negligible nonlocal effect since $\v_s$ 
is perpendicular to $\q$. This result suggests that the best chance for experimentalists
to see the nonlinear Meissner effect is in (110) surface geometry. We again caution, 
however, that for this geometry, we do not take account of the anomalous Meissner currents 
\cite{walter} carried by  the Andreev surface bound state \cite{fogelstrom}.
In Fig. 3 we show the $\theta$ dependence of the normalized penetration depth correction
at $T=0$.

\begin{figure}[h]
\begin{picture}(200,180)
\leavevmode\centering\includegraphics{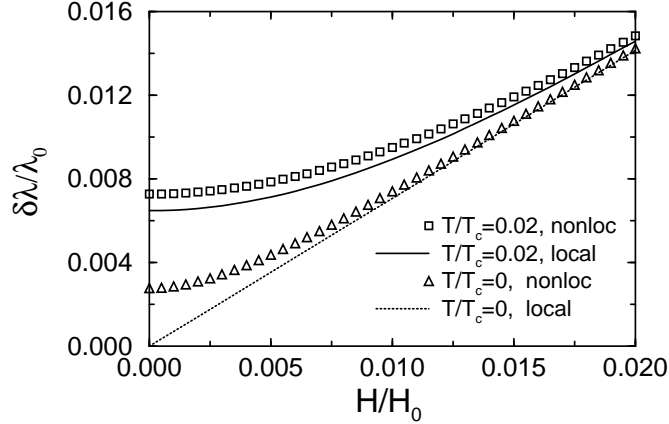}
\end{picture}
\caption{Magnetic field dependence of the normalized penetration depth correction
$\delta\lambda(H,T)/\lambda_0$ for $\theta=0$. $\kappa=100$ is assumed and hence 
$H_{c1}/H_0=0.01$.}
\end{figure}

\begin{figure}[h]
\begin{picture}(200,180)
\leavevmode\centering\includegraphics{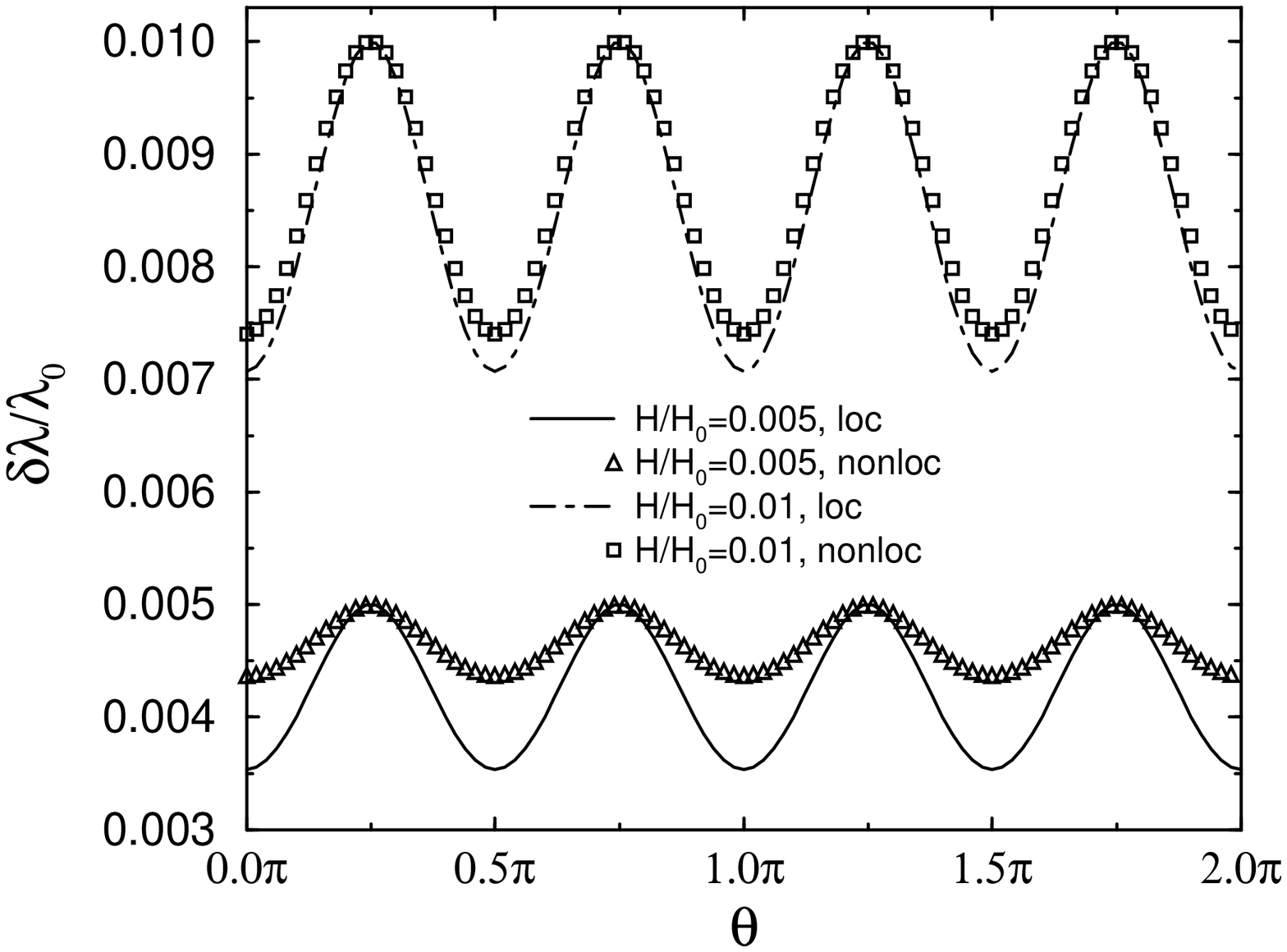}
\end{picture}
\caption{$\theta$ dependence of the normalized penetration depth correction
$\delta\lambda(H)/\lambda_0$ for $T=0$. $\kappa$ is the same as in Fig. 2.}
\end{figure}

Examining the $T$ dependence of $\delta\lambda$ in Eqs. (\ref{pd8}) and 
(\ref{pd9}) we find that for both $h_{\theta l} \gg 1$ and $\ll 1$
the leading $T$ term is quadratic. 
In Fig. 4(a), we plot the $T$ dependence of the penetration depth at fixed $H/H_0$
to show explicitly this feature. We also display $[\lambda(H,T)-\lambda(0,T)]/\lambda_0$
in Fig. 4(b) as a function of $T$ in comparison. Note in particular that 
the magnitude of the field dependence {\it decreases} with increasing temperature, 
as is to be expected for any effect which depends on the sharpness of the $d$-wave nodes. 
 While the size of this decrease is diminished by the nonlocal corrections, there
is never an {\it increase} in field dependence with increasing temperature,
as observed in experiment (see discussion below).

\begin{figure}[h]
\begin{picture}(200,180)
\leavevmode\centering\includegraphics{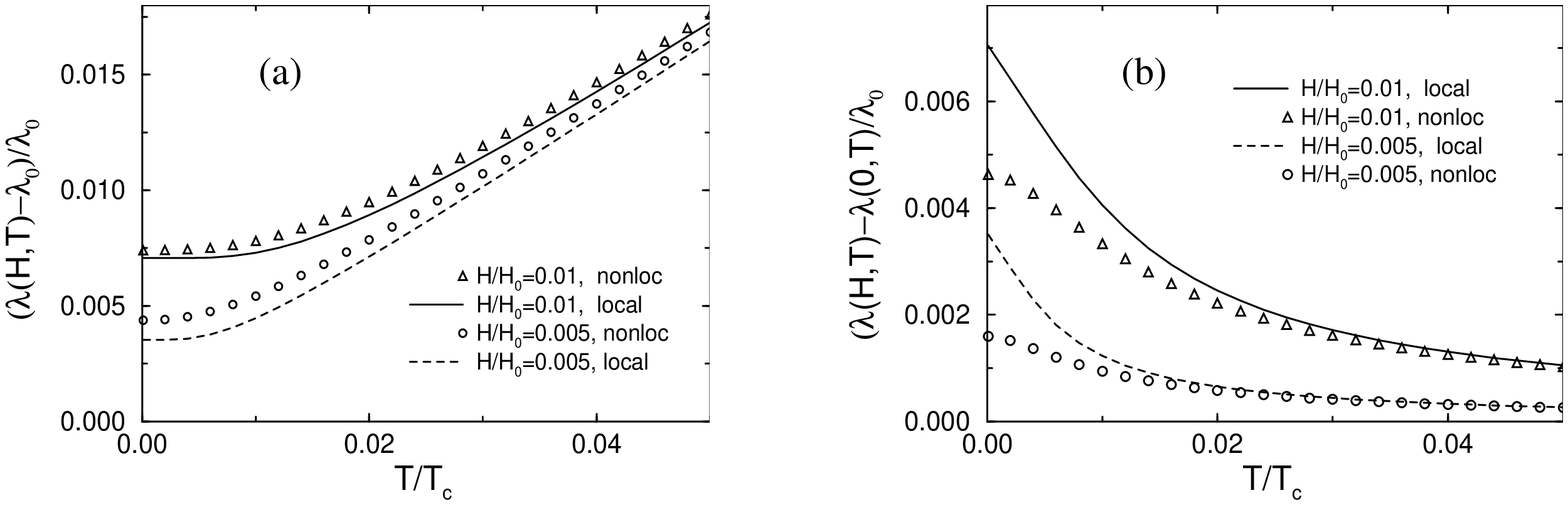}
 \end{picture}
\caption{$T$ dependence of (a) the normalized penetration depth correction  
$\delta\lambda(H,T)/\lambda_0$;
(b)  the normalized penetration depth correction  $(\lambda(H,T)-\lambda(0,T))/\lambda_0$. 
$\kappa$ is the same as in Fig. 2.}
\end{figure}

\subsection{Penetration depth in (001) surface case}
Up to now, we have discussed in detail the penetration depth
when the field is along the $\hat{c}$-axis direction. In that case,
all the interesting physics is in the 2D $ab$ plane and 
it is implied that the possible interlayer transport kinetic energy 
is absorbed into the normal state energy spectrum. 
However, one may also consider the special situation in the quasi-2D 
cuprates in which 
the experimental configuration is characterized by a (001) surface with 
$\H\parallel 
ab$. In this case, $\A$ and $\v_s$ are still in the $ab$ plane (perpendicular to
$H$) forming an angle $\theta$ with the $\hat{a}$ axis. The nonlinear effect that has 
been discussed in the previous sections remains the same. However, the normal
of the surface, and hence the penetration direction $\q$, is perpendicular
to the $ab$ plane. If the quasiparticles are confined rigorously
to the $ab$ plane, they will not contribute to the nonlocal effects. 
But on the other hand, the magnetic field in this case is not screened 
($\lambda\rightarrow\infty$) \cite{GK}.
In a real system, an interlayer coupling exists, leading to 
a nonvanishing $E_{nonloc}^{(ab)}$ \cite{LHWcomment}. Here we confine ourselves 
to the situation of coherent transport along the $\hat{c}$ direction.
This may be a reasonable model at least for the YBCO material. 
To avoid confusing  notation, we keep $\k$ as a 2D
vector in the $ab$ plane, while we use $k_c$ as the momentum along the 
$\hat{c}$ direction. The energy spectrum of the quasiparticles in the superconducting state
in the absence of the magnetic field is $E'_{\k,k_c}=\sqrt{\xi'^2_{\k,k_c}+
\Delta^2_\k}$, where
\begin{eqnarray}
\xi'_{\k,k_c}={k^2\over 2m}+{k^2_c\over 2M}-\mu
\end{eqnarray}  
with $M\gg m$ the effective mass along the $\hat{c}$ direction. 
We have assumed that the pairing order parameter has no $k_c$ dependence.
Now a parallel calculation for the penetration depth can be done. It 
is shown that the momentum $q$ picks out $k_c$ so as to replace
the first argument of the scaling function $F_\lambda$ in Eq. (\ref{rf4})
by ${qk_{Fc}/(2MT)}$. Thus the nonlocal 
energy scale becomes $E^{(ab)}_{nonloc}\simeq {k_{Fc}/(M\lambda_0)}
\simeq \xi_{0c}\Delta_0/\lambda_0$ which is 
no longer $\theta$ dependent. One immediately sees that the main
results presented for $H\parallel c$ case remain:
the available nonlocal effects in (001) surface case serve to cut off
the linear $T$ singularity and modify the $T$ dependence of the penetration
depth to a quadratic one; at extremely low $T$, they cut off the linear $H$
dependence when the field is below a crossover critical field $H^{*(ab)}$
which is found to be $H^{*(ab)}\simeq (\xi_{0c}/\xi_0) H^*$. $H^{*(ab)}$
is much smaller than $H^*$ since the  $\hat{c}$-axis coherence length is 
$\xi_{0c}\simeq 3 A$ as opposed to the in-plane coherence length of 
$\xi_{0}\simeq 15 A$. On the other hand, the lower critical 
field is also much smaller for this geometry, 
$H_{c1}^{(ab)}\simeq {\Phi_0\over 4\pi\lambda_0\lambda_{0c}} {\rm ln}
\bar{\kappa} $ \cite{hc1}, where $\lambda_{0c}$ is the penetration depth for 
supercurrents along the $\hat c$ axis, and $\bar{\kappa}=\sqrt{\kappa \kappa_{0c}} $
with  $\kappa_{0c}$ the $\hat{c}$-axis Ginzburg-Landau parameter. 
 Using $\lambda_{0c}~\simeq ~0.5-1\times10^4 A$, 
we find a large crossover field $H^{*(ab)}\simeq H_{c1}^{(ab)}$, 
making it still 
impossible from a practical point of view to extract a linear-$H$ term
\cite{Cooper}. 

For highly anisotropic samples like BSCCO system, the above argument
might not apply. A full treatment of this problem awaits a generally 
accepted theory of 
the (incoherent)
$\hat c$-axis transport in the normal state.  In addition, it is possible that
surface-barrier effects render the field of first flux penetration much larger
than we have estimated, leaving a large range of fields where the nonlocal effects can
be neglected for this geometry.  We discuss this situation below.

\section{Discussion}

 We first summarize our results. 
In this paper, we have investigated a clean (quasi-) 2D $d_{x^2-y^2}$-wave 
superconductor in the Meissner state based on the weak-coupling  
BCS theory. The existence of nodes on the Fermi surface 
leads to several important effects
at low energy that have nontrivial consequences on the 
free energy and the penetration depth. In addition to the thermal 
excitations of quasiparticles with typical energy scale $T$,  
there are also nonlocal effects, with energy scale $E_{\rm nonloc}$,
due to the divergent effective size of Cooper pairs along nodal directions, 
and nonlinear effects with energy scale $E_{\rm nonlin}$, arising 
from the magnetic field-induced quasiparticle excitations. 
Taking the approximation of a slow varying supercurrent in space, we 
have formulated the Helmholtz free 
energy for the description of Bogoliubov quasiparticles with Doppler-shifted
energy spectrum, and of the response of these quasiparticles to a 
weakly spatially varying magnetic field. 
The free energy is shown, after a Legendre transformation,
 to reduce to Schopohl-Dolgov's free energy
at $E_{\rm nonlin}\ll T$, and to Volovik's singular free energy form
at  $E_{\rm nonlin}\gg T$. The resulting response function includes both 
nonlocal and nonlinear effects,
from which a two-parameter scaling function of the penetration depth $\delta\lambda
(H;T)\simeq 
F(E_{nonlin}/T,E_{nonloc}/T)$ is obtained. The well-known linear $T$
dependence of
$\delta\lambda$ is  obtained only at $T\gg E_{\rm nonlin}, E_{\rm nonloc}$, 
and will be renormalized to $T^2$ whenever
$E_{\rm nonloc}$ or $E_{\rm nonlin}$ is larger than $T$. 
The linear $H$ dependence of $\delta\lambda$ predicted by Yip and
Sauls, the socalled nonlinear Meissner effect, is recovered 
for $E_{\rm nonlin}\gg T, E_{\rm nonloc}$, but is typically changed to 
$H^2$ if  $E_{\rm nonlin}$ is smaller than either $T$ 
or $E_{\rm nonloc}$. 
At extremely low $T$, the nonlocal effects cutoff the nonlinear effect
at $E_{\rm nonlin}<E_{\rm nonloc}$. This happens when $H<H^\ast$. 
Both $E_{\rm nonlin}$ and $E_{\rm nonloc}$ turn out to be geometry
dependent. When the magnetic field is along the $\hat{c}$ axis and the angle $\theta$
that the supercurrent makes with the $\hat{a}$ axis is not near a nodal value,
$H^*\simeq H_{c1}$, leading to the unobservability of the nonlinear Meissner 
effect. When $\theta$ is near a nodal value, $E_{nonloc}$ is so small that 
the nonlinear Meissner effect may be recovered. 
However the effects on $\lambda(H,T)$ of Andreev bound states, neglected in this
work, need to be better understood before this conclusion can be taken seriously. 
\vskip .2cm
Are any of the above predictions supported by experiment?  While the
linear-$T$ temperature dependence in the penetration depth measurements above 1 K in clean
samples of cuprate superconductors has been observed for several years, nonlinear and 
nonlocal consequences of the simple $d$-wave theory of electromagnetic response
in the Meissner state have proven more difficult to confirm.  In zero external field,
any low-$T$ $\delta \lambda\sim T^2$ behavior observed until now occurs in an
experimental situation where the result can more plausibly be attributed
to impurities.  Recent high-resolution resonant coil measurements
have indeed identified linear field dependences $\delta\lambda\simeq bH$ with
coefficients $b$ at temperatures $T\simeq 1$ K not too different from the predictions
of the local, nonlinear theory.  However, the temperature dependence in such cases
is opposite to that predicted, and the linear field dependence is not always observed.
This suggests that earlier reports of the observation of the nonlinear Meissner
effect displaying larger linear-$H$ terms in lower resolution experiments must be treated
with skepticism.  Other manifestations of the  nonlinear Meissner effect, e.g., the ac
transverse magnetization measurements  of Bhattacharya {\it et al.} \cite{Goldmanprivate} have also
failed to observe the predicted angular harmonics,\cite{ZuticValls,ZV} and it is tempting
to conclude that a single mechanism is responsible for the null result obtained by all
types of experiments.
\vskip .2cm
What is the nature of this mechanism?  Our proposal that the linear field dependence
must generically be supressed at fields $H\simeq H_{c1}$ clearly does {\it not}
provide a complete explanation for the failure to observe the nonlinear Meissner
effect, as no $H^2$ dependence has been reported in any sample.  There appear to
us to be two classes of explanation for the existing data on $\delta\lambda(H)$
and transverse magnetization.   First, as suggested by Carrington {\it et al.}, 
\cite{CarringtonGiannetta}
a few vortices trapped in the surface layer can give rise to a ``spurious"
contribution to $\delta\lambda(H)$ due to the microwave excitations of these vortices
in their pinning potentials (Campbell pinning penetration depth).
\cite{Campbellpendepth}  Depending
on the size of the pinning force, this can give a result, $\propto H$, of roughly 
the correct magnitude.
It is not clear whether the transverse magnetization experiments can
be quantitatively explained by such a contribution. 
\vskip .2cm
The second possible explanation for the disagreement of the experiments
with both the local and nonlocal $d$-wave theories was put forward by 
Bhattacharya {\it et al.}, \cite{Goldmanprivate}
 who suggested the existence of a small {\it bulk}
subdominant order parameter component, e.g., of $s$ symmetry, which condenses
in a state $\pi/2$ out of phase with the dominant $d$ component.  Such a 
state has a true ``minigap" of order the size of the $s$ component.  The
authors of Ref. \cite{Goldmanprivate} estimated that a minigap of a few degrees $K$ would
be sufficient to eliminate the nonlinear signal in their experiments on
YBCO.  In such a scenario, any experiment which depends on the existence 
of quasiparticles
well below the minigap scale must report a null result.  Several other
experiments {\it have} reported significant quasiparticle densities
well below this temperature.  In particular, the thermal conductivity
measurements of Taillefer et al. \cite{tailleferuniversal}
 reveal the existence of a normal-fluid density in both the 
 YBCO and BSCCO systems down to at least
50 mK.  We therefore do not believe this proposal is tenable.
\vskip .2cm 
An argument against the importance of nonlocal effects in recent experiments
with $H\parallel ab$ which may provide a clue to the 
origins of the spurious signal has been given recently by Bhattacharya et
al.\cite{Goldmanprivatecomment} These authors point out that the field of first flux entry,
$H_{c1}^*\simeq 250$-$300$ Oe, appears to be an order of magnitude larger than the
Ginzburg-Landau lower critical field
$H_{c1}\simeq 30$-$50$ Oe, of order the crossover field between nonlocal and local field
dependence for 
YBCO in this geometry.  There should therefore, according to these authors, be a
rather large intermediate field region before flux penetrates the sample where the
local theory applies and the nonlinear Meissner effect should be observable. 
This argument, while intriguing, begs the question of why the observed field of first
flux penetration is so different from the Ginzburg-Landau  $H_{c1}$ \cite{LHWreply}. The obvious explanation
is a geometrical surface barrier effect of the type discussed by Bean and Livingston,
and by De Gennes.\cite{barrier}  Vortices which nucleate in the surface layer in the
intermediate field range, while thermodynamically stable in the bulk, are expelled by an
image potential due to the surface. 
The intermediate  field ``Meissner" state is therefore thermodynamically unstable, and
vortices may be easily trapped at the sample corners or within the skin depth. 
 Two recent high-resolution
penetration depth measurements \cite{CarringtonGiannetta,Bidinosti}
failed to measure the
predicted intrinsic (nonlinear) temperature dependence in this field range,
but did  observe a large linear-$H$ field dependence
most likely attributable to trapped vortices. 
 On the other hand, the lack of hysteresis observed by Bhattacharya {\it et al}.
when cycling above $H_{c1}^*$ argues against a significant number of pinned vortices.
Nevertheless, the thermodynamic instability of the Meissner state suggests that
the existence of a region of perfectly laminar surface flow with $H_{c1}<H<H_{c1}^*$
where the simple 
theories presented above apply may be too naive.  The correct picture of the
surface  layer in such a situation may be one of fluctuating supercurrents whose time
averaged effect on  those quasiparticles present may be similar to that of trapped
vortices.  These fluctuations would make the field  penetration layer 
instantaneously highly nonuniform, potentially leading to a broadening 
and a redistribution of spectral weight of the 
predicted transverse magnetization harmonics.\cite{ZuticValls} 
We believe the only field regime where
one can safely assume a true Meissner response is for fields 
below the thermodynamic $H_{c1}$, i.e., in the nonlocal 
response regime.

\vskip .2cm
We propose two types of experiments which could help clarify the
present impasse regarding theoretical interpretation of 
experimental results.  The first is a direct measurement of
the magnetization using a dc superconducting quantum interference device.
   This has not been yet
performed as a function of applied magnetic field, to our knowledge.  If
the requisite sensitivity can be obtained, such an experiment would
have the advantage of eliminating the Campbell penetration depth
contributions, which may be dominating the signal in the resonant
coil experiments.\cite{CarringtonGiannetta,Bidinosti}  The second
experiment, to maximize the possiblity of seeing nonlocal effects and 
eliminate the uncertainties in the analysis with $H||ab$, would be a 
measurement with field $H||c$ on a long cylindrical sample with geometrical
axis coinciding with the crystalline $\hat{c}$ axis.  These may help to
settle this long-standing controversy.

\acknowledgements{   The authors are grateful to A. Carrington, F. Evers, 
M. Fogelstr\"om, M. Franz, R. Giannetta, A. Goldman, S. Kirchner, J. Kroha, 
N. Schopohl, and Y.-J. Wang for helpful communications.  Partial support was 
provided by the A. v. Humboldt Foundation, NSF Grant No. DMR-9974396, and the Deutsche
Forschungsgemeinschaft.} 

\section*{Appendix: Renormalization of the penetration depth}

In Sec. IV, we calculated the penetration depth in the presence of a constant $v_s$. 
In a real system, however, the $v_s$ field
of the screening current is varying in space.
The correction to the penetration depth due to the spatially varying nature
of $v_s$ needs to be examined in order to check the validity of the 
qualitative conclusions we have reached. Note that
the rigorous result can only be obtained through 
the solution of the full self-consistent magnetostatic problem, 
from which analytical information is difficult to obtain. However, 
we can improve our previous results by employing better
approximations. We may imagine the system to be subdivided 
into many layers, in each of which $v_s$ is roughly constant. But $v_s$
in different layers is not the same. 
The response function in  real space is now approximatedly defined as
\begin{eqnarray}
\K(y,y')\simeq\K \left [y-y',v_s\left ({y+y'\over 2}\right )\right ] .
\label{rfr1}
\end{eqnarray} 
 The only terms not included in this approximation involve gradients of $v_s$.
However, in the regime where $E_{nonlin}\gg E_{nonloc}$, we 
estimate ${\bf\nabla}\v_s/m\ll v^2_s$ and hence gradients of $v_s$ are 
negligible where the nonlinear effect is important at all.

According to $\A=(mc/e)\v_s$, Eq. (\ref{ampere}) can be rewritten in 
an operator representation
\begin{eqnarray}
v_s=-{4\pi e\over mc^2} \,
\left (\nabla^2-{4\pi\over c}\hat{\K} \right )^{-1}j_{ext}\, . 
\label{vsr1}
\end{eqnarray}  
Using the expansion 
\begin{eqnarray}
{4\pi\over c}\hat{\K}\simeq \lambda^{-2}_0+{4\pi\over c}\delta\hat{\K} \,,
\label{rfr2}
\end{eqnarray}  
we get
\begin{eqnarray}
v_s\simeq -{4\pi e\over mc^2} \,
\left [(\nabla^2- \lambda^{-2}_0)^{-1}j_{ext}
+(\nabla^2- \lambda^{-2}_0)^{-1} \left ({4\pi\over c}\delta\hat{\K} \right )
(\nabla^2- \lambda^{-2}_0)^{-1}j_{ext} \right ] \, . 
\label{vsr2}
\end{eqnarray}  
In the coordinate representation $v_s$ takes the form of
\begin{eqnarray}
v_s(y)\simeq v^{(0)}_s(y)+v^{(1)}_s(y) , \label{vsr3}
\end{eqnarray}  
\begin{eqnarray}
&&v^{(0)}_s(y)=-{eH\over mc}\,
\lambda_0e^{-|y|/\lambda_0} \label{vsr4} \\
&& v^{(1)}_s(y)=-{8eH\over mc^2} \, \int^\infty_{-\infty}
{dq_1\over 2\pi}{e^{iq_1y}\over q^2_1+\lambda^{-2}_0}
\int^\infty_{-\infty}dR \int^\infty_{-\infty} dq_2
{e^{i2R(q_2-q_1)}\over (2q_2-q_1)^2+\lambda^{-2}_0}\delta\K(\q_2,\v_s(R))
\, .
\label{vsr5}
\end{eqnarray}  
At this stage, the $\q$ and $\v_s$ dependence of $\delta\K$ can be approximately 
taken to be the same as what we got in Sec. II, with the point kept in mind that $v_s$ 
is a function of space which needs to be solved self-consistently based on Eqs.
(\ref{vsr3})-(\ref{vsr5}). 
Using the definition in Eq. (\ref{pd1}) and the relation 
$H(y)=(mc/e)(dv_s(y)/dy)$,  we find
\begin{eqnarray}
\lambda_{spec}=-{mc\over eH}v_s(0), 
\end{eqnarray}
which turns out to be
\begin{eqnarray}
\lambda_{spec}\simeq\lambda_0-{8\over c}
\int^\infty_{-\infty}
{dq_1\over 2\pi}{1\over q^2_1+\lambda^{-2}_0}
\int^\infty_{-\infty}dR \int^\infty_{-\infty} dq_2
{e^{i2R(q_2-q_1)}\over (2q_2-q_1)^2+\lambda^{-2}_0}\delta\K\left (\q_2,\v_s(R) \right ) .
\label{pdr1}
\end{eqnarray}  
It is easy to check that in the case of a constant $v_s$, 
the penetration depth in Eq. (\ref{pdr1})
reduces exactly to that in Eq. (\ref{pd2}).

Now we insert Eqs. (\ref{rf4}) and (\ref{scalingf2}) into Eq. (\ref{pdr1}) 
to obtain the normalized change in the penetration depth
\begin{eqnarray}
{\delta\lambda_{spec}\over \lambda_0}={T\over \pi\lambda_0\Delta_0}
\sum_{l=\pm 1}u^2_{\theta l}
\int^\infty_{-\infty}{dt_1\over 2\pi}\int^\infty_{-\infty} dR
{I_{\theta l} (t_1,R)\over t^2_1+1}, \label{pdr2}
\end{eqnarray}
\begin{eqnarray}
&&I_{\theta l}(t_1,R)= \int^{2|h_{\theta l}(R)|}_0 dt_2
\left [{e^{i2(t_2-t_1)R/\lambda_0}\over (2t_2-t_1)^2+1}+
{e^{i2(t_2+t_1)R/\lambda_0}\over (2t_2+t_1)^2+1} \right ]
\beta_{\theta l} |v_s(R)| \nonumber\\
&&\;\;\;\; +\int^\infty_{2|h_{\theta l}(R)|} dt_2
\left [{e^{i2(t_2-t_1)R/\lambda_0}\over (2t_2-t_1)^2+1}+{e^{i2(t_2+t_1)
R/\lambda_0}\over (2t_2+t_1)^2+1} \right ] \,
\left [{\pi\over 4}{a_{\theta l}t_2\over \lambda_0}+\beta_{\theta l} |v_s(R)|\lambda^2_0
{\beta^2_{\theta l} v_s^2(R)+\pi^2\over 6a^2_{\theta l}t^2_2} \right ] ,  \label{ii}
\end{eqnarray}  
where $a_{\theta l}=v_Fu_{\theta,-l}/(2\sqrt{2}T)$ and 
$\beta_{\theta l}=mv_Fu_{\theta l}/(\sqrt{2}T)$.
Clearly the nonlinear and the nonlocal corrections
to the penetration depth are involved in $I_{\theta l}(t_1,R)$. 

For $h\gg 1$
it is straightforward to find 
that the leading correction to the penetration 
depth is linear in $H$. It is given by the first integral in $I_{\theta l}(t_1,R)$ 
with $v_s(R)$ replaced by $v^{(0)}_s(R)$ and
the upper bound of the integral $2|h_{\theta l}(R)|$ expanded to $\infty$.
After proceeding with this one gets
\begin{eqnarray}
{\delta\lambda_{spec}\over \lambda_0}
\simeq {\zeta_\theta\over \sqrt{2}}{H\over H_0}=
{\pi\zeta_\theta\over 3\sqrt{2}}\kappa^{-1}h \,, 
\;\;\; \;\; h_{\theta l}\gg 1 \,, \;\;\;\;\; l=\pm 1 .  \label{pdr3}
\end{eqnarray}  
Comparing with the first term in Eq. (\ref{pdlocal}) we find that
an extra prefactor $2/3$ is acquired through this renormalization scheme.
When we compare this result with Yip-Sauls result (\ref{pdys}), however, 
as we mentioned at the end of Sec. IV A, we note a different definition for the 
penetration depth is used in their paper, and an extra prefactor 2 should be
expected. This is to say that if we use the definition of Yip-Sauls for the penetration depth, 
we should find it 2 times as large as that in Eq. (\ref{pdr3}). The reason that 
the expected 2 is actually missing here is again the consequence of the 
perturbation treatment of the coupling of the quasiparticles to the $\A_{\q\neq 0}$
mode, the detailed discussion of which has been presented in Sec. III.

For $h\ll 1$, the leading correction to the penetration depth 
is the contribution from the nonlocal effects and is independent 
of $H$. It can be obtained from the second integral in $I_{\theta l}(t_1, R)$
by setting the lower bound $2h_{\theta l}(R)$ to be $0$.
The leading $H$-dependent correction to the penetration depth is found
to be quadratic of $H$. After some algebra we get
\begin{eqnarray}
{\delta\lambda_{spec}\over \lambda_0}
\simeq {\pi\over 8\sqrt{2}}\sum_{l=\pm 1}d_{\theta l1}\kappa^{-1}+
c_1\lambda_0h^2\kappa^{-1}\sum_{l=\pm 1}
{u^4_{\theta l}\over u_{\theta,-l}} ,
\;\;\;\; \;\; h_{\theta l}\ll 1 \, , \;\;\;\; l=\pm 1 .  \label{pdr5}
\end{eqnarray}  

We find that the first term, i.e., the correction from the nonlocal effects 
is exactly the same as the first term in Eq. (\ref{pd9}). This means, as 
expected, that the nonlocal effects are not influenced by the spatial 
varying nature of $v_s$. The second term coincides qualitatively with 
that in  Eq. (\ref{pd9}), but acquires an insignificant extra prefactor through the 
renormalization scheme.  If this expansion were continued, it is clear the result
obtained would converge to the result of Yip and Sauls\cite{YipSauls} in the local
limit.

}


\begin{references}
\bibitem{note1} Usually there is an additional term representing onsite
repulsion term that prevents an usual $s$-wave pairing. This term 
is important in studying the stability problem of a $d$-wave superconductor. 
But in this work we assume there is always $d$-wave pairing thus decoupling this 
term by a trivial number operator that can be simply absorbed into the 
chemical potential. Besides, throughout this paper, we choose $k_B=\hbar=1$ 
units.
\bibitem{hc1} The expression for $H_{c1}$ given here is 
a direct result of the anisotropic Ginzburg-Landau theory and is 
more accurate than that
appearing in our paper (Ref. \cite{LHWPRL}).
\end{references}
\end{document}